\documentclass[lettersize,journal]{IEEEtran}
\usepackage{graphicx}  
\usepackage{caption} 
\usepackage{amsmath,amsfonts}
\usepackage{algorithmic}
\usepackage{algorithm}
\usepackage{array}
\usepackage[caption=false,font=normalsize,labelfont=sf,textfont=sf]{subfig}
\usepackage{textcomp}
\usepackage{stfloats}
\usepackage{url}
\usepackage{verbatim}
\usepackage{graphicx}
\usepackage{cite}
\begin{document}

\title{Network Digital Twin for Congestion-Aware Predictive Traffic Routing using Graph MPNNs}

\author{Umer Iqbal,~\IEEEmembership{Member,~IEEE}, Ashiq Anjum,~\IEEEmembership{Member,~IEEE},  Anthony S Conway,~\IEEEmembership{Member,~IEEE}, Mathias Kern,~\IEEEmembership{Member,~IEEE} , Anasol Pena Rios,~\IEEEmembership{Senior Member,~IEEE}}




\maketitle

\begin{abstract}
Telecom networks scale with growing users and data-intensive applications, generating heavy traffic that causes congestion, reducing throughput, increasing delay, and raising computational costs. Traditional routing protocols act only after performance degradation, making them unsuitable for dynamic traffic and topological changes. Addressing these challenges requires a routing approach that adapts in real time, scales with network growth, operates without disrupting active services, and provides continuous feedback for congestion-aware traffic optimisation. The Network Digital Twin (NDT) addresses these needs by mirroring global network behaviour using Message Passing Neural Networks (MPNNs) through bidirectional communication with the physical network. To align the NDT with physical network behaviour, synthetic traffic is generated with increasing load across topological structures that incrementally scale as routers are added. These topologies are created by graph-generating models such as Erdős–Rényi, Barabási–Albert, and Watts-Strogatz, customised with vertex degree limitations. The NDT collects performance metrics from routers and links, and MPNNs classify edges based on local vertex and global network behaviours. Based on these classifications, feedback is sent as Policy-Based Routing (PBR) protocol commands to each router, enabling optimal traffic distribution across links of the physical network. A comparative analysis of Multiprotocol Label Switching (MPLS) configured with requirement-specific Open Shortest Path First (OSPF) is conducted using performance metrics to evaluate the effectiveness of the proposed approach in optimising network traffic. Experimental results show that the increased file transfer rate by 180.5\%  and a reduction in delay, congestion, and computational cost by 51.89\%, 73.36\% and 40.98\% respectively, demonstrating that the approach is predictive, adaptable, and scalable for growing traffic needs telecommunication network.

\end{abstract}


\section{Introduction}
As technology advances, digital infrastructures host more data-driven applications and services such as cloud platforms, streaming, artificial intelligence, B5G networks \cite{9509294,9144301},and IoT \cite{8972389,ullah2022new,maddikunta2022industry}. These services depend heavily on telecommunication networks\cite{8961984,9237460}, and because they are data intensive, they generate increasing volumes of traffic. Along with this, the number of users also grows, which needs the networks to be scaled or enhanced to meet demand. Scaling involves adding and removing the routers or adapting to random connectivity within a topological structure. This expansion distributes traffic along multiple routing paths, where some of which become congested while others remain uncongested, depending on the routing protocol and consumer activity on the network. Such dynamic traffic and topological changes due to technological advances  place significant pressure on the network, leading to congestion, a decrease in throughput, increased delay, and higher computational costs \cite{umer1}. Conventional routing protocols \cite{6994333} react only after traffic performance degradation occurs, leaving them unsuitable for dynamic network environments where continual changes in traffic patterns and topology further disrupt performances. Addressing these challenges requires solutions that adapt in real time, scale with growth, operate without disrupting live services\cite{kiruthiga2025evolution}, and provide continuous feedback for congestion-aware optimisation. The Network Digital Twin (NDT) meets these needs by replicating network behaviour through Message Passing Neural Networks (MPNNs), maintaining bidirectional communication with the physical network, and improving traffic performance while preserving service quality\cite{alashjaee2024optimizing,9540090}.

\begin{figure}
    \centering
    \includegraphics[width=1\linewidth]{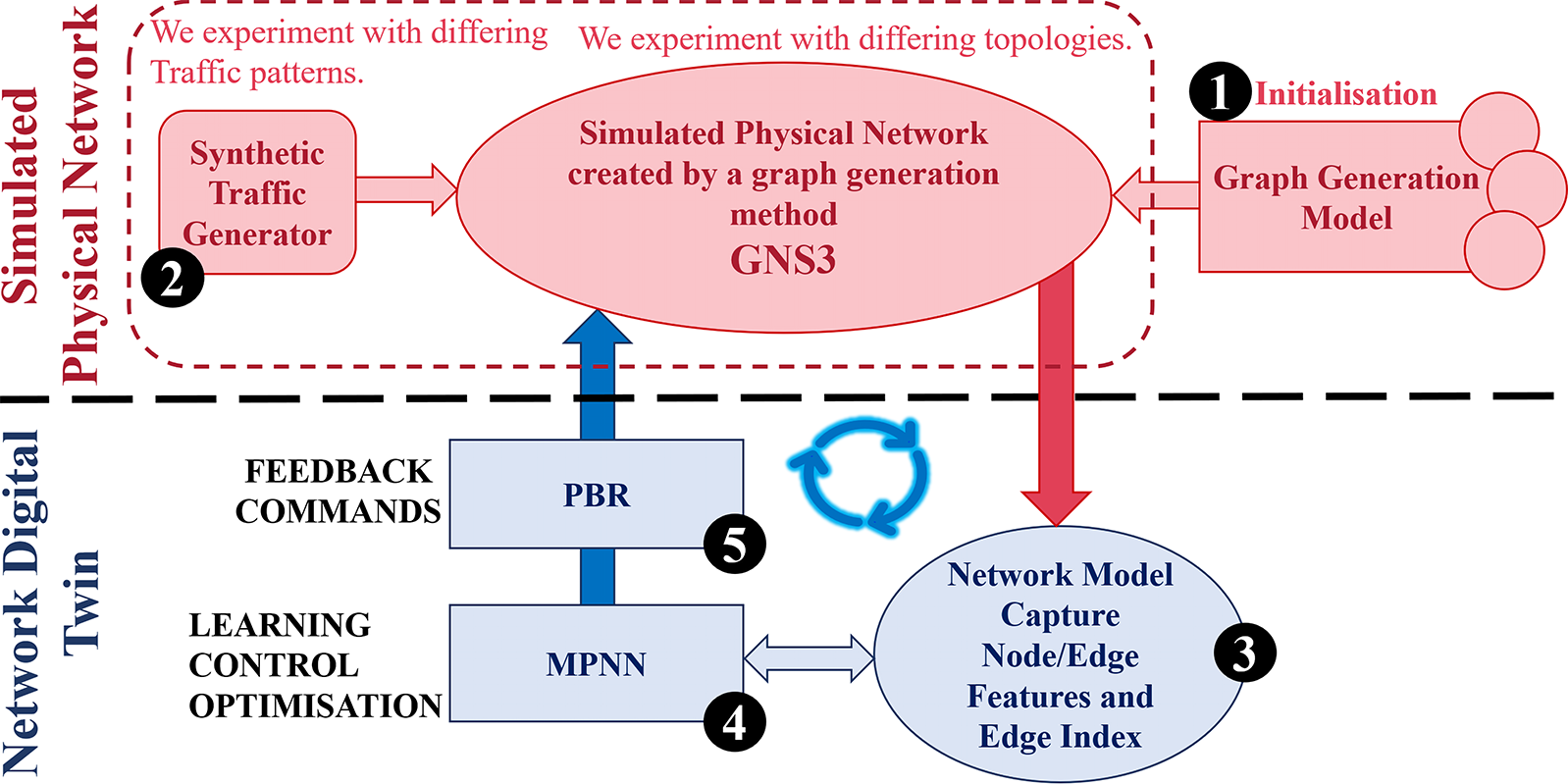}
                          
 \caption{Bidirectional Communication in the Network Digital Twin (NDT)}
     \label{fig:DBNDT}
       \vspace{-2.0em}
\end{figure}

The NDT \cite{9795043} emerges as a promising solution: a virtual replica of the physical network that observes, learns, and responds in near real time. By maintaining bidirectional communication as shown in fig. \ref{fig:DBNDT} with the physical network, the NDT mirrors overall network traffic behaviour to know how dynamic traffic impacts the traffic and delivers proactive feedback for congestion-aware traffic optimisation on the physical network. The behavior replicated by the NDT is achieved by first retrieving information from all edges connected to each vertex in the network. This information forms the data model, which is used to derive the vertex features and edge indices, enabling the model to adapt effectively to changing network and traffic conditions. The second component of the NDT employs MPNNs to process this data model and classify edge congestion at each vertex \cite{prabowo2023message}, integrating local vertex characteristics with the global network behavior. The NDT then uses these edge classifications to generate MPNN-based PBR (Policy-Based Routing) \cite{agbon2024ai} commands, which intelligently reroute \cite{9855684} the traffic \cite{umoga2024exploring} from congested to uncongested edges of each vertex in the network. To make this behaviour adaptable, diverse topological structures are explored, and for scalability, topologies of different sizes are tested as user numbers and traffic loads increase. On the physical side, traffic is generated to bring these scenarios to life and evaluate performance under realistic conditions. 

There is a limitation for physical networks in adaptability and scalability to test the above scenarios because it is difficult to constenantly modify them, so experiments have been carried out in simulation. To examine adaptability, diverse topologies are constructed to demonstrate how dynamic traffic reshapes traffic performance. The traffic is generated with progressively heavier loads by increasing file size, and the network is scaled up by adding new routers, which reflect the physical network. These topologies are generated mathematically using graph-generation models such as Erdős–Rényi, Barabási–Albert, and Watts–Strogatz \cite{drobyshevskiy2019random,van2024random}, by applying probabilistic or deterministic rules to design flexible and adaptive structures. Each model contributes distinct properties: Erdős–Rényi captures uniform randomness \cite{9815224,10364868}, Barabási–Albert builds hubs through preferential attachment \cite{barabasi1999emergence,chakrabarti2012preferential}, and Watts–Strogatz combines clustering with short paths to represent small world effects \cite{watts1998collective,li2024improved}.Together, these models provide adaptable foundations, refined with vertex degree constraints to reflect router capacities and ensure practical telecom implementation. By contrast, models such as Random Geometry Graphs (distance-based and unsuitable for MPLS), Kronecker Graphs (highly complex), and Hyperbolic Graphs (better suited for social networks) do not align with telecom requirements.

As customised topologies are deployed in simulation, which makes traffic becomes essential to drive realistic scenarios. In simulation, traffic patterns can be controlled, while in physical networks traffic depends on unpredictable user behaviour. Generating traffic  patterns on real infrastructure would be complex and costly, making synthetic traffic the practical alternative. The traffic flows on diverse topologies of different structures and sizes, making the NDT generalisable, adaptable, and scalable. Dynamic flows are generated from multiple user hosts by gradually increasing file sizes after each file transfer is completed, steadily raising the traffic load. This growth allows evaluation of performance and congestion control strategies under realistic conditions. Traffic flowing through varied topologies enables the NDT to capture overall network behaviour. Packets are collected from edges, and the extracted edge index, vertex, and edge features are transformed into feedback, delivered as PBR commands. PBR is selected because it is more flexible than static IP routing: it enables redistribution based on traffic type, packet size, or IP address, while static IP routing relies solely on IP address. This flexibility makes PBR particularly effective for congestion-aware traffic optimisation on networks.

Conventional routing protocols intervene only after networks begin to degrade, keeping them reactive rather than proactive. In dynamic environments where routers and links are continually added, removed, or reconfigured, this delay proves costly. MPLS improves traffic engineering\cite{shawl2014review}\cite{ridwan2020recent} compared with traditional methods. While MPLS forwards packets via Label-Switched Paths (LSPs) and, under MPLS-TE \cite{bernardez2023magnneto}, can steer traffic via constraint-based routing with resource designation (e.g., bandwidth, QoS), it typically reacts to present network situations (e.g., resource availability or link failure) instead of consistently predicting future traffic load, limiting it's flexibility for truly traffic-aware optimisation. So, it shows that it cannot continuously predict and adjust to changes, leaving gaps in adaptability and scalability. MPLS networks can be configured with interior gateway protocols (IGPs) such as RIP (Routing Information Protocol), EIGRP (Enhanced Interior Gateway Routing Protocol), OSPF (Open Shortest Path First), and IS-IS (Intermediate System to Intermediate System).When selecting an IGP, one needs to make sure that the requirements are satisfied so that the routing configuration can be set up for traffic flow in the P(Provider) routers topologies that are generated using graph-based models. OSPF \cite{kohler2003mpls} meets these requirements better than other IGPs like RIP and EIGRP, which are better for edge routers. IS-IS is good for very large networks but not as beneficial for smaller simulation networks, so OSPF is the most effective and beneficial choice here. To demonstrate the advantages of the suggested method, metrics such as throughput, delay, and congestion for NDT routing utilising MPNN-based PBR are compared to MPLS configuration with OSPF on the physical network.

This work makes several contributions. First, it integrates the NDT with MPNNs, creating a system that enables congestion-aware traffic optimisation while also learning network behaviour and predicting how traffic can be improved on physical infrastructure. Second, to start activity traffic is generated across mathematically generated topologies of different structures and sizes. By gradually increasing the traffic load on these structures, the study ensures evaluations are generalisable, scalable, and adaptable, making optimisation strategies realistic and applicable to physical networks. Third, it demonstrates the advantages of MPNN-based PBR as a programmable protocol, turning the network into a programmable network capable of flexibly redistributing traffic in ways static routing cannot. Finally, the approach is validated by comparing the proposed MPNN-based PBR framework against MPLS configured with OSPF, showing measurable improvements in delay, congestion, computational cost, and throughput. These results confirm the predictive, adaptive, and scalable nature of the NDT.

The research contributed to the NDT by integrating modified graph generation models with MPNN-based PBR to create an implementable and optimisable traffic routing on the physical network. These contributions help in  the development of adaptive, efficient, and scalable telecom networks that can meet the demands of next-generation applications. The remainder of this paper is structured as follows: Section \ref{sec:Literature_Review} reviews related work. Section \ref{sec:NDT} gives the details about the NDT. Section \ref{sec:methodology} outlines the methodology, including graph generation, network configuration, traffic generation, information retrieval, MPNN, and traffic rerouting on vertex. Section \ref{sec:implementation} details the implementation of the proposed NDT. Section \ref{sec:experiment} describes the experimental setup and evaluation process. Section \ref{sec:results} presents the results and discusses their implications in the context of telecom network optimisation. Finally, Section \ref{sec:conclusion}  highlights future work directions and concludes with key insights and summarises the contributions of this study.

\section{Literature Review}

\label{sec:Literature_Review}

Advancement in graph-based network topology design, traffic engineering, and machine learning-driven optimisation gives the foundation for adaptive and scalable telecommunications networks. These approaches are designed to optimise throughput while minimising congestion and delays, thereby enabling real-time traffic management in dynamic network environments.

\subsection{Graph-Based Network Topology Design}

Graphs are essential for systematically creating topologies and quantitatively analysing telecom networks. There are different methods for this, such as standard, algorithmic graph generation models and learning-based ones. Standard methods \cite{lim2016review} are traditionally used for creating topologies like buses, rings, stars, meshes, and trees, which are simple and easy to implement. Algorithmic models like Erdős–Rényi (ER), Barabási–Albert (BA), and Watts–Strogatz (WS) more precisely represent the complex network system. Learning-based methods \cite{ZHOU202057}, such as graph neural networks (GNNs), initially require the data to learn and predict the graph structure. Algorithmic graph generation models are better because they don't require the datasets for training and predicting graph structure. On the other hand  the  computation resources and the standard methods are predefined structures, which are unable to capture the complexities. Algorithmic graph generation models offer valuable insights into complex networks without any dataset and can also be modified to be practically implementable for scenarios like vertex degree limits or limiting network size. In this situation, where we need to analyse the complex network with no initial data, only a few parameters are needed;  the algorithmic graph generation models are more suitable for a realistic and efficient solution.

\subsection{MPLS Traffic Flow and Routing}

MPLS employs label-based forwarding for traffic routing \cite{awduche2002internet}, optimising latency and bandwidth allocation \cite{soorki2014label, alwayn2001advanced}. It models a network using vertices to represent routers and edges to define the connections between them. Research demonstrates MPLS's effectiveness in reducing latency, optimising bandwidth allocation \cite{mustapha2019intelligent}, and enabling dynamic routing in telecom infrastructures \cite{1510723}.

MPLS enhances traffic flow between interconnected LANs through WANs by prioritising optimal path selection. However, traditional MPLS relies on the reactive network configuration that based on resource constraints and network events, limits it ability for adaptive and predictive to real-time traffic load conditions. This research integrates machine learning-based routing to dynamically adjust paths, ensuring efficient load balancing.

\subsection{Machine Learning in Network Traffic Management}

Graph Neural Networks (GNNs) have demonstrated effectiveness in modelling network topology \cite{liang2022survey} and traffic flow \cite{tam2022graph, yan2024dampnn}. Within this category, Message Passing Neural Networks (MPNNs) \cite{prabowo2023message,10704927} provide fine-grained traffic control by leveraging vertex-edge relationships.

MPNNs update network state by continuously analysing traffic conditions, allowing vertices to make real-time adjustments to optimise flow. Unlike traditional algorithms, MPNNs dynamically redistribute traffic based on congestion levels, latency, and bandwidth utilisation \cite{zhang2024graph}. Adaptive traffic management is used in this study using MPNNs and PBR strategies to make changes in real time as network conditions change.

\subsection{Network Digital Twin for Telecom Networks}

Network digital twins replicate the behaviour of networks from real-world scenarios, enabling predictive optimisation \cite{ferriol2022building, ngo2023empowering}. Previous applications have focused on managing congestion via reinforcement learning \cite{david2021inference}, optimising resource allocation in 5G networks \cite{8764584}, and enhancing the reliability of IoT networks \cite{vaezi2022digital}.

Even with these improvements, current network digital twins rely significantly on physical testbeds, which makes it hard to scale and automate. This research discusses a modular network digital twin that enables traffic generation, topology design, and virtual network configuration independent of real infrastructures, hence assuring adaptable and scalable network experimentation.

\subsection{Conventional vs AI-Based Routing}
The shift from conventional to AI-based bidirectional routing signifies the continued development of intelligent traffic management systems. Researchers first proved that just replacing conventional interior gateway protocols like RIP or OSPF with a software-defined network (SDN) controller. For instance, on a highly congested IoT network, throughput increased from 0.078 Gbps to 0.087 Gbps, representing an 11.7\% \cite{sharma2023novel} improvement, while end-to-end delay reduced by around 19\%\cite{sharma2023novel}. During periods of low traffic, SDN just assisted a reduction in delay (about 27\%) without impacting throughput. During periods of high traffic, the advantages were notably minimal, with throughput seeing little growth and delay reduced by about 1.86\% \cite{sharma2023novel}.Following that, more advanced AI-based methodologies evolved. A deep graph convolutional-reinforcement learning method enhanced throughput by 7.8\% and reduced delay by 16.1\% in comparison to OSPF \cite{abrol2024deep}. A deep reinforcement-learning protocol for secure SDN-IoT networks (DQQS)  as compared to OSPF  enhanced throughput from 8 Mbps to 14.5 Mbps (about an 81\% increase) and reduced delay from 88 ms to 52 ms (approximately a 41\% reduction) \cite{arif2024dqqs}. The African Vulture metaheuristic and other algorithms outperformed deep reinforcement learning by 16.9\% and classical routing by 71.8\% \cite{chen2024dynamic}. A multi-agent deep deterministic policy gradient approach, conversely, enhanced throughput by around 3\% and reduced delay by roughly 7\% \cite{kuang2025rs}. Advanced congestion-control systems that transmit switch-to-switch signals achieved flow completion times 1.2 to 2 times faster \cite{li2024congestion}, and novel SDN load-balancing designs asserted a 70\% increase in efficiency for load distribution \cite{ahmad2022effectively}. The bidirectional NDT uses the MPNN-based PBR approach, which has increased file-transfer speeds by 180.5\% and reduced delay by 51.89\%, congestion by 73.36\%, and computational cost by 40.98\%.  This makes the NDT both a diagnostic and prescriptive tool, bridging the gap between simulation and implementation.
\section{Network Digital Twin}
\label{sec:NDT}
The Network Digital Twin (NDT) is a virtual counterpart of the physical network that mirrors its behaviour and responds in near real time. Unlike static models, the NDT continuously observes and learns from the live physical network, adapting to dynamic traffic conditions and evolving topologies. Its effectiveness comes from bidirectional communication: the physical network sends real-time metrics such as throughput, delay, and congestion to the NDT, where the physical network is mirrored through a data model that captures vertex and edge features representing its operational state. This data model is then processed and used for learning through Message Passing Neural Networks (MPNNs), enabling the system to understand network behaviour and predict optimal configurations. The learning outcomes are subsequently mapped into PBR commands, which are sent as feedback to the physical network to achieve adaptive and efficient traffic routing at vertex and network level. In this way, the NDT captures both local router-level and global network behaviours, enabling predictive analysis and congestion-aware optimisation. It also provides a safe and flexible simulation environment, where dynamic traffic can be generated, diverse topologies can be tested, and PBR commands can be validated before being applied to the physical network. This makes the NDT both a diagnostic and prescriptive tool, bridging the gap between simulation and implementation.

\subsection{Research Gaps and Objectives}
Even though a lot of research has been done on network design and traffic management, there are still some significant challenges that need to be solved, much as the ones mentioned in the abstract and introduction. First, many current graph generation models don't take consideration of practical boundaries such vertex degree limits, which makes them less  applicable in telecom networks where router capacity is bound. Second,  even though programmable networks can dynamically reconfigure the routers to optimize traffic flow, but present implementations don't fully exploit this feature to optimize network traffic flow in real time. Real-time changes in traffic demand mean that forwarding devices need to be updated quickly and data collecting has to be very accurate. But for now, programmable network infrastructures face challenges by delayed state propagation, limited hardware flexibility, and delayed monitoring. Third, a lot of network digital twins still rely on physical testbeds a lot and don't use simulations well to optimise in real time. Fourth, the problems of dynamic traffic growth and changing topology structures over time are not adequately addressed in current approaches, limiting their ability to cope with evolving network states. These limitations leave telecom networks lacking adaptability, scalability, and predictive optimisation—the very requirements emphasised earlier.
Building on these gaps, this research aims to:
\begin{enumerate}

    \item Design graph generation models that incorporate real-world constraints, such as vertex degree limits, to create practical and scalable telecom topologies.

   \item Employ MPNNs and PBR to enable adaptive traffic management that proactively adjusts to dynamic network conditions, making the network itself programmable and establishing PBR as a programmable protocol.

   \item Develop a modular NDT that integrates traffic generation, topology design, and simualtion network configuration, reducing dependence on costly physical testbeds while supporting bidirectional communication in near real-time for traffic optimisation on physical networks.
\end{enumerate}

\section{Methodology}
\label{sec:methodology}

\begin{figure}
    \centering
    \includegraphics[width=1\linewidth]{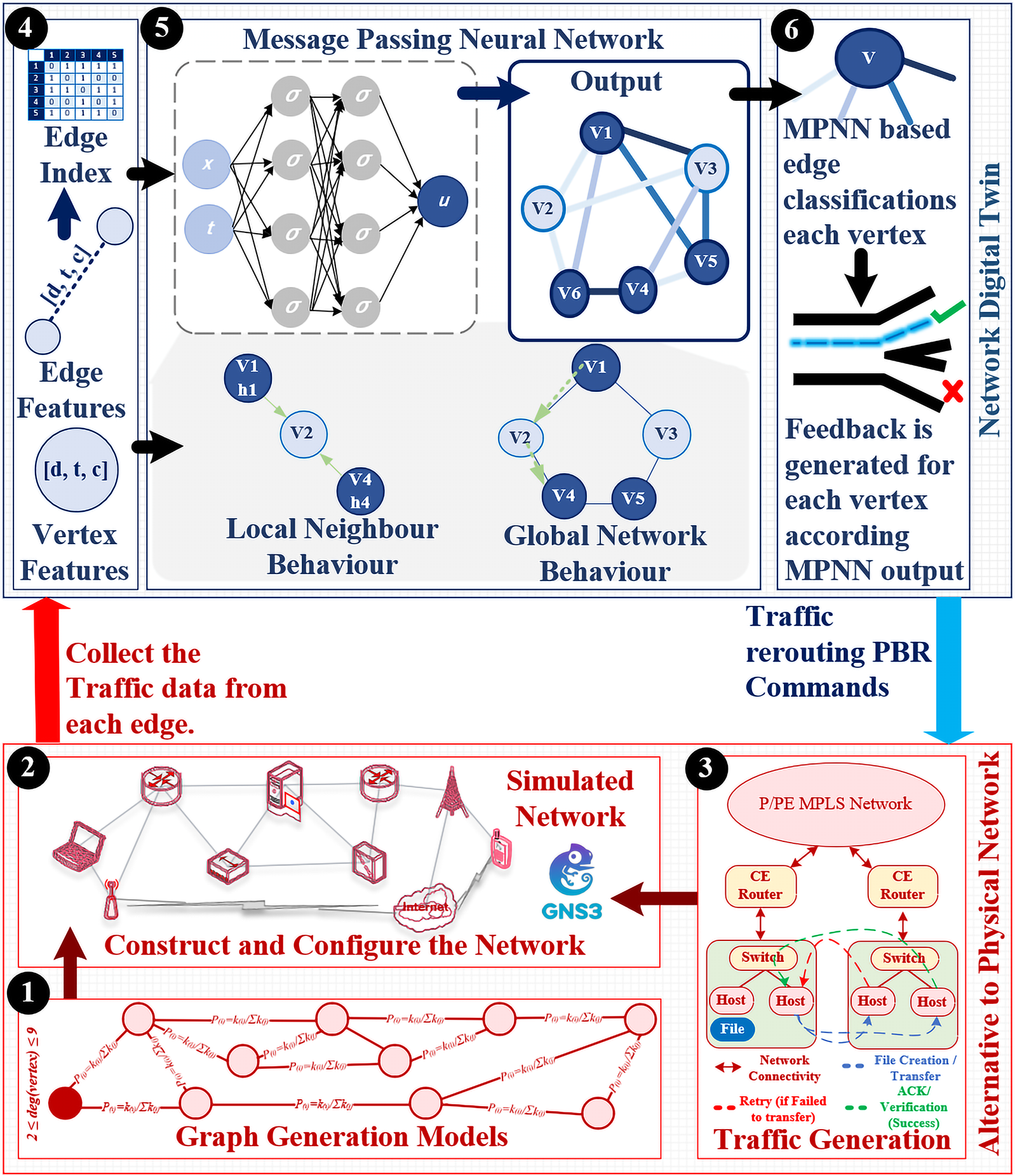}

           \caption{Architecture of the Network Digital Twim (NDT): Graph Generation models construct the network, traffic is generated to initiate activities, and these activities are captured in one direction as input to the NDT, which processes MPNNs to provide feedback in the other direction for optimal traffic reroutes on the network.}
    \label{fig:NDT}
    \vspace{-1.5em}
\end{figure}

The Network Digital Twin\textbf{NDT}\cite{10847826} as in fig. \ref{fig:NDT}, is formalised in this study for \textbf{near real-time, programmable network traffic optimisation} \cite{10365500} that scales and adapts to dynamic traffic and topological changes of physical networks. It starts with the graph generation models  to  generate \textbf{graph topologies}, because the NDT requires realistic network topological structures to capture connectivity and traffic dynamics before optimisation can be performed. Since the physical network is created in a simulation environment, topological structures are essential for representing its behaviour, and they provide the flexibility needed for adaptability. These topologies scale  with the addition of new routers, allowing the NDT to evaluate network traffic performance by changes in connectivity and scalability of topologies. To maintain realism, the topologies are generated with the constraint of \textbf{vertex degree limitations} \( \Delta(v) \) to have realistic network structural connectivity \( \kappa(G) \) that reflects the finite number of interfaces in real routers. This constraint guarantees that the generated topologies are both mathematically consistent and practically implementable. Within these topologies, each vertex \( v \in V \) and edge  \( e \in E \)  is represented by feature mappings \( F(v), F(e) \in \mathbb{R}^3 \), capturing delay, throughput, and congestion, while the edge index \( E_i \subseteq V \times V \) encodes connectivity between vertices. The NDT uses these features with the help of MPNNs to classify the congestion states of links attached to each vertex by knowing local vertex and global network behaviour. This link classification helps the NDT decide a percentage of traffic rerouting from congestion to the uncongested link of each vertex in the network. These rerouting decisions of NDT are sent back to each router in the physical network as PBR commands, completing the cycle of bidirectional communication. Through this continuous process of observation, learning, and feedback, the NDT optimises traffic flow, reduces congestion on edges or links, and balances traffic among the edges or links, providing a scalable and adaptive solution for congestion-aware traffic optimisation in physical networks.

\subsection{Graph-Based Topology Design}
To construct and configure the network in simulation, the topology is generated using graph generation models, such as \textbf{Erdős–Rényi}, \textbf{Barabási–Albert}, and \textbf{Watts-Strogatz}. To ensure realistic applicability in telecom networks that reflect a finite number of interfaces on real routers of the network, a degree constraint $2 \leq \deg(v) \leq 9$ is imposed, where $\deg(v)$ represents the degree of vertex $v$. The minimum degree of 2 guarantees full connectivity by avoiding isolated or stub vertices, while the maximum degree of 9 reflects the practical interface limitations of routers. This constraint preserves the theoretical integrity of the models while ensuring their suitability for real-world network simulations.

\subsubsection{Erdős–Rényi Model}
The Erdős–Rényi model generates a random graph \( G(n, p) \), where \( n \) represents the number of vertices and random probability \( p \) represents the chances whether an edge exists or not (it is either 1 or 0) between any two vertices independently. For every pair of vertices \((u,v)\), a random draw decides whether the edge is added \((u,v) \in E\) or not, ensuring that all edges are chosen independently. Each vertex maintains at least 2 and at most 9 connections, ensuring compliance with the vertex degree constraint. Since edge formation is independent for each vertex pair, the probability-based process governs the structure while preventing isolated or overly connected vertices. 

\subsubsection{Barabási–Albert Model}
The Barabási–Albert (BA) model with probability \( p \) is based on the degree of the current vertex. This means that vertices with more linkages are more likely to get new connections. The model used to scaling up of the network with the addition of every new vertex, needs to connect with existing vertices and make their vertex degree higher in network. This is reflecting the growth of telecommunication networks where core routers accumulate more connections over time, thereby producing realistic scale-free structures. To make it practically implementable, a vertex degree limitation is used to have a minimum of 2 and a maximum of 9 vertex degrees.  This retains the scale-free property while ensuring the topology remains consistent with real-world router interface limitations.

\subsubsection{Watts-Strogatz Model}
The Watts-Strogatz model creates a graph by first creating a ring lattice of \( V \) vertices to make sure that all local connections are the uniform before adding randomisation to get the small-world effect, with each vertex connected to its \( k \) closest neighbours. Each edge \( e \) is subsequently rewired with a probability \( p \), adding stochasticity while preserving local clustering, a hallmark of small-world networks. For practical use, the model limits the degree of each vertex to between 2 and 9. A minimum degree of 2 helps mitigate isolation and poor connectivity, which causes vertices to come together to form a single structure. A maximum degree of 9 prevents too many connections from happening, which would lead to redundancy and inefficiency. This controlled degree distribution keeps the network resilient, making both the structure and the way it works more effective for real-world uses.

\subsubsection{Comparing models for generating graphs in simulated networks}

GNS3 simulation creates the generated network topologies, which are then mapped into an MPLS network configuration that includes both Wide Area Network (WAN) and Local Area Network (LAN) environments.  The WAN consists of \textbf{Provider (P)} and \textbf{Provider Edge (PE)} routers, while the LAN comprises \textbf{Customer Edge (CE)} routers, \textbf{switches}, and \textbf{hosts (Linux servers)}.\textbf{P (Provider (Core))} routers are represented as the subset \( V_P \subseteq V \) within the network graph \( G = (V, E) \). These routers provide packet forwarding in WAN networks. The topology \( V_P \) is directly derived from the produced graph \( G \), assuring consistency with WAN-level setups.

Table \ref{tab:T1} is comparing the topologies generated through generation models like Erdős–Rényi (ER), Barabási–Albert (BA), and Watts-Strogatz (WS). The comparison is across topologies with 5, 10, and 15 provider (P) routers, with their number of edges, number of paths, and average hop path length. As the number of P routers increases, all methods show a rise in the number of edges and paths. BA, which forms networks through preferential attachment, consistently generates a moderate number of paths and edges with relatively low average path lengths in the network. In contrast, ER produces the highest number of paths as compared to the other two topologies, leading to longer average path lengths with an increase in the number of routers. WS, known for its clustering and short path characteristics, generally results in fewer paths but with longer average hop lengths compared to Barabási–Albert. 
\begin{table}[ht]
\centering
\begin{tabular}{|l |l|l |l |l |l|}
\hline
\textbf{P Routers} & 
\textbf{Topology} & 
\textbf{Edges} & 
\textbf{Paths} & 
\textbf{Path Length} \\
\hline
5& BARABASI ALBERT  & 7 & 4 & 2.25 \\
 & ERDOS RENYI  & 6 & 3 & 2.33 \\
 & WATTS STROGATZ  & 6 & 6 & 3 \\\hline
10& BARABASI ALBERT  & 17 & 34 & 3.12 \\
 & ERDOS RENYI  & 19 & 353 & 6.37 \\
 & WATTS STROGATZ  & 11 & 9 & 5.22 \\\hline
15& BARABASI ALBERT  & 27 & 408 & 5.6 \\
 & ERDOS RENYI  & 29 & 5804 & 9.77 \\
 & WATTS STROGATZ  & 17 & 31 & 7.58 \\ \hline
\end{tabular}
\vspace{0.1em}
\caption{Graph Generation Models: Topology Summary}
\vspace{-2.0em}
    \label{tab:T1}
\end{table}

\subsection{Dynamic Traffic Generation}

To initiate the activities in the simulated network, dynamic traffic is generated through automatic file transfers between hosts of different LANs across the network by increasing the file size with every iteration. The iterative nature of this approach enables adaptive refinement of traffic patterns, providing a quantitative way for evaluating network performance and congestion control strategies in simulated environments.

For file creation (traffic modelling), the file size is \( F_n \), which starts at \( F_0 = S_{\text{min}} \) and goes up to \( n \) iterations to create network traffic. Each time the file size increases by \(\Delta F\), it makes the network volume go up according to the stepwise function below:

\[
    F_n = F_{n-1} + \Delta F, \quad \Delta F = 140428 \text{ bytes}
\]

where \( F_n \) is the file size at iteration \( n \) and \( \Delta F \) is the constant increase in file size. This formulation ensures a controlled and systematic escalation of network load.

Data is transmitted between the hosts of different LANs using the \texttt{rsync}  is a unix-utility over SSH, ensuring secure transmission. And on every iteration old files are removed and new file are created with different sizes. Given a source vertex \( S \) and a destination vertex \( D \), the transfer rate is defined as:

\[
    R_t = \frac{F_n}{T_t}
\]

Where \( R_t \) the transmission rate (bytes per second) and \( T_t \) is for the observing file transfer time. This rate is continuously logged to assess network throughput efficiency.

\textbf{Failure Detection and Recovery (Fault Tolerance)}
If a transfer fails, the system triggers an adaptive recovery mechanism, restarting the host or resending the file with the same size. Traffic generation operates through concurrent tasks, modelled as an event-driven system. Each transmission task \( T_i \) has the below operations.

\[
    T_i = \{O_1, O_2, \dots, O_m\}
\]

where each operation \( O_j \) includes file transfers, vertex status verification, and error handling. Execution follows a state transition model:

\[
    S : \{\text{Running, Executing, Completed, Failed}\}
\]

If \( S \) is set to \textit{Running}, this means no action is needed. If it is set \textit{executing}, then it needs to be check the the source host for file transfer progress. If it is set to \textit{completed}, this mean it need to check all  vertices to complete their transfers in that iteration. If it set to\textit{Failed},  then it needs check failure cause and then takes action. If source or destination host failure, the system restarts the affected host and resend the files. If the failure is due to a port issue, the port is reactivated before attempting to resend the file. This systematic approach guarantees constant traffic generation and strong failure recovery.

Transfer events are logged with timestamps, source-destination pairs, file sizes, and completion status. The data is aggregated into a structured dataset for further statistical modelling, including file transfer rate and reliability.

\subsection{Information Retrieval from Simulation Network}

The Information Retrieval Module is an important part of the NDT. It is  designed to extract and analyse near real-time data from the network environment. The vertex and edge features are extracted from this information. These features are then served as input baselines for congestion-aware network traffic optimisation using MPNN-based PBR  in the network.

The number of features is extracted from information retrieval from the network, where each is represented mathematically to provide a structured way for analysis. Key features include:
\textbf{Delay ($D$):} The edge delay (D) is the inter-arrival time, which is the temporal interval between the arrival of two successive packets. Mathematically, if the arrival times \( T\) of two successive packets are the  \( n \)th  packet at \( T_n \) and \( (n+1) \)the  packet at \( T_{n+1} \), then:

\[
D = T_{n+1} - T_n
\]

The vertex delay is calculated as the difference between the first packet arrival time on an ingress edge, $t_{\text{min ingress arrival}}$, and the last packet arrival time on an egress edge, $t_{\text{max egress arrival}}$, and is given by:

        \[
        D = t_{\text{min ingress arrival}} - t_{\text{max egress arrival}},
        \]

The throughput for the edge is the total size of data($S$) transmitted on the edge in the time interval $\Delta t$, calculated as $t_{\text{max arrival}} - t_{\text{min arrival}}$. Throughput is defined as 
    \[
    T = \frac{S}{\Delta t},
    \]
        throughput for the vertex, where $S$ the total data size $D$ is the delay on the vertex. It is calculated as
        \[
        T = \frac{S}{D},
        \]

The congestion on an edge and vertex $T$ is the throughput, and $T_{\text{max}}$ is the maximum possible throughput during network traffic.
    \[
    C = \frac{T}{T_{\text{max}}} \times 100,
    \]

The edge data size is the cumulative volume of traffic data transmitted during a time interval, where it $s_i$ represents the size of each packet in bytes.
    \[
    S = \sum_{i=1}^n s_i,
    \]
  
    The total size of data processed by a vertex during a time interval is the total data size from ingress $S_{\text{ingress}, k}$ and egress $S_{\text{egress}}$ edges connected to the vertex. 
        \[
        S = \sum_{k=1}^n \left(S_{\text{ingress}, k} + S_{\text{egress}, k}\right),
        \]

A computational cost  \( W_e \) of each network edge  \( e \) is calculated by knowing the \( D_e \) delay or time consumed to process the data and the congestion metric on edge \( C_e \). Both  \( D_e \) \( C_e \) are weighted equally at 0.5 because the more congestion, the more time consumed and packet loss, which results in more computation needed. The  \( D_e \) and \( C_e \) are two different metrics with different ranges of values, so we need to apply the min-max normalisation to scale them with a uniform range [0, 1] to avoid bias from the difference in their original ranges.
\[
W_e = 0.5 \cdot D_e  + 0.5 \cdot C_e
\]
The total computational cost for the entire network is then calculated as the sum of individual edge costs:

\[
\text{Total Cost} = \sum_{e \in E} W_e 
\]

By capturing comprehensive and accurate performance data, the Information Retrieval Module ensures that the Network Digital Twin operates effectively, enabling real-time, adaptive network management.

\subsection{MPNN-Based Traffic Optimization}

\begin{figure}
    \centering
    \includegraphics[width=1\linewidth]{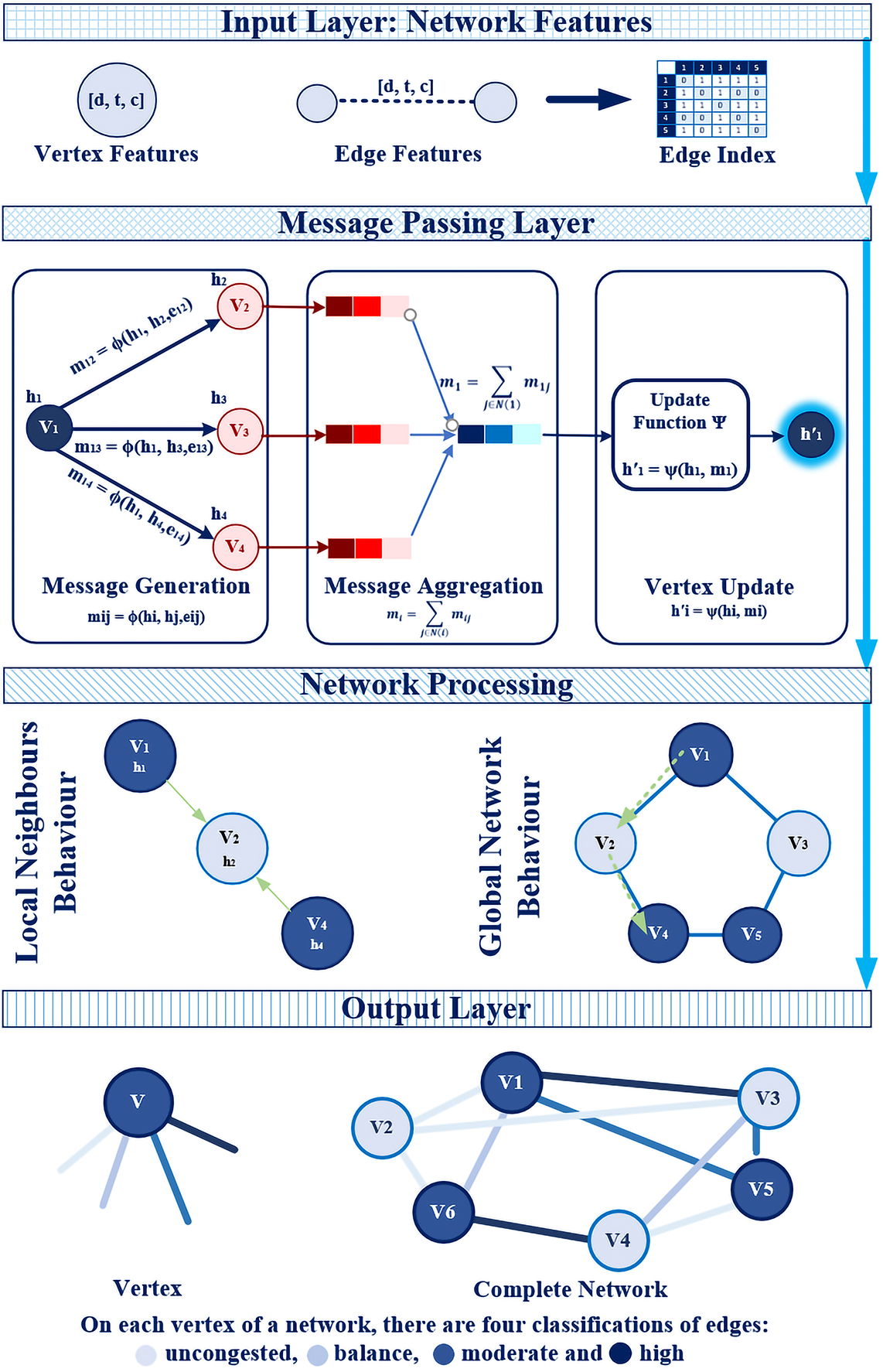}

                        \caption{Four-layer MPNN architecture: input, message passing, network processing, and output classifying edges per vertex for traffic rerouting. }
    \label{fig:MPNN}
    \vspace{-1.5em}
\end{figure}

The NDT uses MPNNs, shown in fig. \ref{fig:MPNN}, to optimise traffic flow in dynamic networks by classifying edges of each vertex in the network. The MPNN iteratively processes edge and vertex features, along with their edge index, to classify edge congestion. During each iteration, it aggregates information from neighbouring vertices to refine congestion classification and improve decision-making. By analysing the topological relationships within vertices and edges, the model identifies congested edges and calculates an adaptive traffic rerouting function to manage the traffic flow in near real time. This improves the network traffic performance in different traffic patterns and network topologies and sizes.

The inputs of MPNN for model training consist of edge index, vertex, and edge features. These inputs enable the model to adapt to changes in network topology and traffic conditions, ensuring scalability and generalisation.

Each vertex \( v_i \in V \) is characterised by a feature vector \( \mathbf{x}_v \in \mathbb{R}^3 \), given by:
\[
\mathbf{x}_v = [ \text{Delay}(v_i), \text{Congestion}(v_i), \text{Throughput}(v_i) ]
\]
These attributes represent the vertex’s processing delay, current congestion, and throughput, respectively. This vector encodes the local behaviour to the model  at each vertex, which helps in \textbf{generalising} it in different network topologies.

Each edge \( e_{ij} \in E \) is described by a feature vector \( \mathbf{x}_e \in \mathbb{R}^3 \), defined as:
\[
\mathbf{x}_e = [ \text{Delay}(e_{ij}), \text{Congestion}(e_{ij}), \text{Throughput}(e_{ij}) ]
\]
These features capture the transmission delay, congestion, and throughput on the edge \( e_{ij} \). These features help the model to be \textbf{adaptable} under different traffic and congestion conditions on each communication link.

The edge index \( E_i \subseteq V \times V \) defines the relationship between vertices, encoding the \textbf{global network topology}. It helps the model to know overall data propagation among the vertices  in the graph. It ensures that the model can scale efficiently as the network grows in size.

The \textbf{message-passing layer} is a key component in the graph-based learning framework, facilitating information exchange between vertices through their connecting edges. This process consists of three stages. In the first stage of \textbf{message generation}, each vertex communicates with its neighbours by message generation based on its own features, the features of the neighbouring vertex, and the edge attributes connecting them. The message  function \( \phi \) is defined as

\[
    m_{ij} = \phi(h_i, h_j, e_{ij})
\]

where \( h_i \) and \( h_j \) represent the vertex feature and \( e_{ij} \) denote the edge feature.
In the second stage of \textbf{message aggregation}, a vertex aggregates incoming messages from all its connected neighbours to capture structural and contextual information. This aggregation is performed as

\[
    m_i = \sum_{j \in N(i)} m_{ij}
\]

where \( N(i) \) represents the neighbouring vertices of vertex \( i \).

Once the messages are aggregated, each \textbf{vertex updates} its state using an update function \( \psi \), which takes into account its previous state and the aggregated message:

\[
    h_i' = \psi(h_i, m_i)
\]

This iterative process enables vertices to refine their feature representations by incorporating both their local information and the influence of their neighbours.

\begin{figure}
    \centering
    \includegraphics[width=1\linewidth]{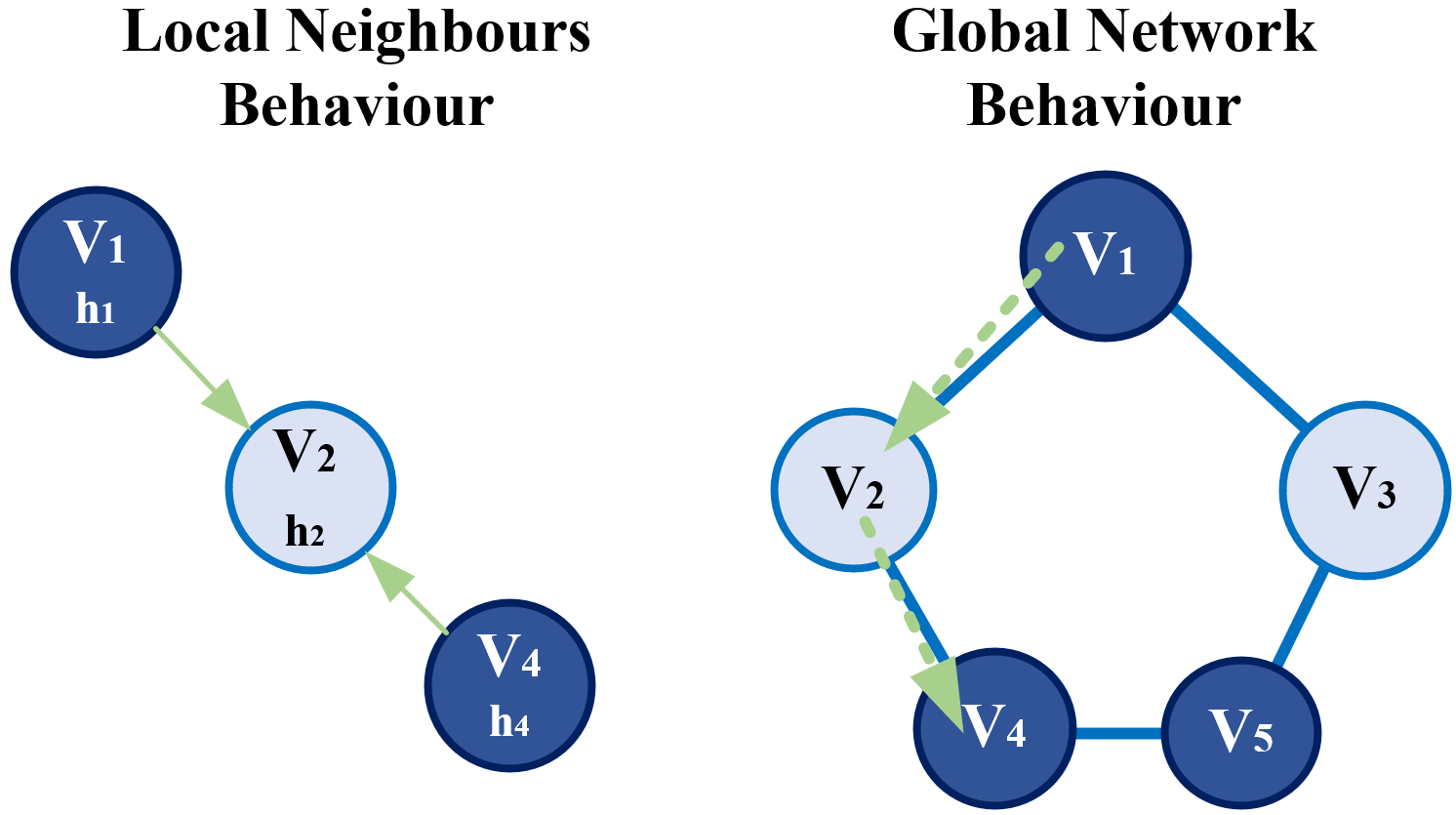}
            \caption{Local vertex behaviour emerges through interactions with neighbours, shaping global network behaviour.}
    \label{fig:LG}
    \vspace{-1.5em}
\end{figure}

The third stage of \textbf{network processing} is done in two levels, as shown in fig. \ref{fig:LG}. The \textbf{Local Neighbours Processing}, where a vertex exchanges information with directly connected neighbours, and \textbf{Global Network Processing}, where information propagates beyond immediate connections.

In the \textbf{local neighbour's behaviour}, each vertex updates its feature representation based on messages received from its neighbours. First, it processes the \textbf{message computation}, where a vertex gathers information from its neighbours. In the diagram, the vertex \(V_2\) receives messages from \(V_1\) and \(V_4\):
  \[
        m_{12} = \phi(h_1, h_2, e_{12}) \quad \in \mathbb{R}^8
    \]
     \[
        m_{42} = \phi(h_4, h_2, e_{42}) \quad \in \mathbb{R}^8
    \]
Secondly, it processes these gathered/received messages to \textbf{ aggregate} them by summing to create a unified message representation.
     \[
        m_2 = m_{12} + m_{42}
    \]
Finally, these aggregated messages are used for \textbf{vertex update}, the vertex updates its state:
    \[
        h_2' = \psi(h_2, m_2) \quad \in \mathbb{R}^8
     \]

Beyond local interactions, information propagates across the network, capturing  \textbf{global network behaviour} through higher-order relationships. In the \textbf{first-order neighbours}, vertices directly exchange information with adjacent vertices. For example:
    \[
        V_1 \rightarrow \{V_2, V_3\}, \quad V_2 \rightarrow \{V_1, V_4\}
    \]
 In the \textbf{second-order neighbours}, a vertex indirectly interacts with further connections. In the diagram:
    \[
        V_1 \rightarrow V_2 \rightarrow V_4, \quad V_3 \rightarrow V_5 \rightarrow V_4
    \]

The \textbf{Feature Propagation} is through multiple iterations; vertices integrate information from their \textbf{k-hop neighbourhood}, where  \( k \) represents the number of layers in the propagation process.

The MPNN model \textbf{outputs congestion classifications for each edge} \( e_{ij} \). These classifications guide the traffic rerouting process by identifying edges that require traffic redistribution.

After passing through the message-passing and aggregation steps, each edge \( e_{ij} \) is classified into one of four congestion levels:
\[
\hat{y}_{ij} \in \{1, 2, 3, 4\}
\]
\textbf{where}: \textbf{1}: Highly Congested, \textbf{2}: Moderately Congested, \textbf{3}: Balanced, and \textbf{4}: Uncongested

The MPNN is \textbf{trained iteratively} on \textbf{diverse network topologies and dynamic traffic scenarios}. These topologies are represented as graphs, where the \textbf{edge index} is determined based on \textbf{vertex and edge features}, ensuring a structured representation of connectivity. During each training iteration, the model updates its parameters based on \textbf{real-time traffic data} and \textbf{predicted congestion levels} for edges connected to each vertex.

\subsection{Training Process of MPNN for Traffic Optimization}

To achieve optimal performance, the model is trained on 10.4 million records over six distinct topologies, obtained from three different graph generation models with two different network sizes.  In the training, the model minimises the \textbf{cross-entropy loss} between the predicted congestion levels \( \hat{y}_{ij} \) and the true congestion labels \( y_{ij} \), which are derived from observed network conditions. The loss function is formulated

\[
\mathcal{L}(\theta) = \sum_{i,j} \mathcal{L}_\text{cross-entropy}(\hat{y}_{ij}, y_{ij})
\]

Then \( \mathcal{L}_\text{cross-entropy} \) calculate the divergence between actual and expected congestion states.

Model parameters \( \theta \) are updated using \textbf{gradient descent}, enabling the MPNN to iteratively refine its congestion-aware traffic optimisation strategy. As training progresses, the model learning is more effective at classifying the edge congestion attached to each vertex in the network.

\subsection{Traffic Rerouting}

Once congestion classifications are obtained from the MPNN, the next step is to compute an optimal traffic rerouting strategy. The objective is to redistribute traffic efficiently, balance congestion, and maximise throughput across network edges. To determine whether traffic rerouting is necessary or not on basis on edge classification on each vertex in network. This is done by calculating the traffic volume that need to rerouted from congested edges to  to underutilised or balanced edges of each vertex, by ensuring efficient load distribution and improved network performance.

\subsubsection{Traffic Rerouting on Vertex}

Let \( E_u \), \( E_b \), \( E_m \), and \( E_h \) represent the number of edges in each congestion category for a vertex, defined as the cardinality of their respective sets: \( E_u = |\text{underutilized}| \), \( E_b = |\text{balanced}| \), \( E_m = |\text{moderate}| \), and \( E_h = |\text{high}| \). For each vertex, congested edges are represented by \( E_c = E_h + E_m \), while the uncongested edges are represented by  \( E_u + E_b \) in the network.

The total number of edges in the network, \( E_{\text{total}} \), is the sum of the edges across all categories:
\[
E_{\text{t}} = E_{\text{u}} + E_{\text{b}} + E_{\text{m}} + E_{\text{h}}
\]

As shown in the equations below, rerouting will not occur if the total number of edges for a vertex is equal to either the number of \textbf{underutilised edges} or the number of \textbf{congested edges}. Additionally, if there are only two edges connected to the vertex, rerouting is not feasible, as there are insufficient alternative paths for traffic rerouting. 
 
\[
E_{\text{t}} = E_{\text{u}} + E_{\text{b}}  \quad \text{or} \quad  E_{\text{t}} = E_{\text{m}} + E_{\text{h}} \quad \text{or} \quad  E_{\text{t}} =  2
\]

Rerouting occurs if both the number of \textbf{underutilised edges} and the \textbf{congested edges} are greater than 1:
\[
E_{\text{u}} + E_{\text{b}} \geq 1 \quad \text{and} \quad E_{\text{m}} + E_{\text{h}} \geq 1
\]

This ensures that there is sufficient capacity on alternative paths to handle the rerouted traffic without causing further congestion. 

\subsubsection{Percentage of Traffic Rerouted}
The rerouting weight determines how much traffic will be rerouted from the congested edges to the uncongested edges. The sum of the weight of highly or moderately weighted edges, using the total number of edges, is denoted as {Totaledges}, which helps in calculating the percentage of traffic to be rerouted on the vertex of the network. This calculation will reroute more traffic from congested edges when they are fewer in number compared to uncongested ones and vice versa. This helps in balancing the traffic between the congested and uncongested edges of the vertex in the network, shown in fig. \ref{fig:TRG}. The weights for moderately and highly congested links are calculated as follows:

\[
    W_{\text{m/h}} = w_{\text{ratio}} \times
    \left( 100 - \left( \frac{E_{(m/h)}}{\text{Totaledges}} \times 100 \right) \right)
\]
\( w_{\text{ratio}} \) is a congestion-based weight factor, set to 1 for highly congested links and 0.75 for moderately congested links.

The percentage of inflow traffic to be redistributed is given by:
\[
    percentage = \frac{W_m + W_h}{E_m + E_h}
\]

To prevent excessive load on underutilised edges and to make the congested edge at the level of balanced edges, we put the limit of 50 for redistribution percentage; this will help the uncongested edges not to be congested and congested links to not move more traffic to make the uncongested edges congested.
\[
    \text{if } percentage > 50, \text{ then } percentage = 50
\]

\subsubsection{Rerouting Strategy}

After calculating the percentage of traffic to be rerouted, the next step is to determine which source and destination IPs are rerouted from congested to uncongested edges of the vertex in the network. This ensures efficient traffic distribution and balances the network load.

To make informed rerouting decisions, historical traffic data is analysed to identify the dominant \textbf{source-destination (S-D) pairs} of IPs that passed through the vertex in the network. The system retrieves past traffic records and calculates the \textbf{packet size percentage} for each S-D pair relative to the total traffic handled by the router:

\[
\text{Traffic Percentage} (T_{sd}) = \frac{\sum_{t=1}^{T} S_{sd}(t)}{\sum_{t=1}^{T} S_{total}(t)} \times 100
\]

where: $S_{sd}(t)$ represents the packet size for the S-D pair $sd$ at time $t$, $S_{total}(t)$ is the total packet size processed by the router at time $t$, $T$ denotes the total observation period.

Once the percentage $P$ of traffic to be rerouted is determined, S-D pairs are sorted in \textbf{ascending order of traffic contribution}. The system iterates through the ranked S-D pairs, summing their traffic percentages until the cumulative total reaches $P$:

\begin{figure}
    \centering
    \includegraphics[width=1\linewidth]{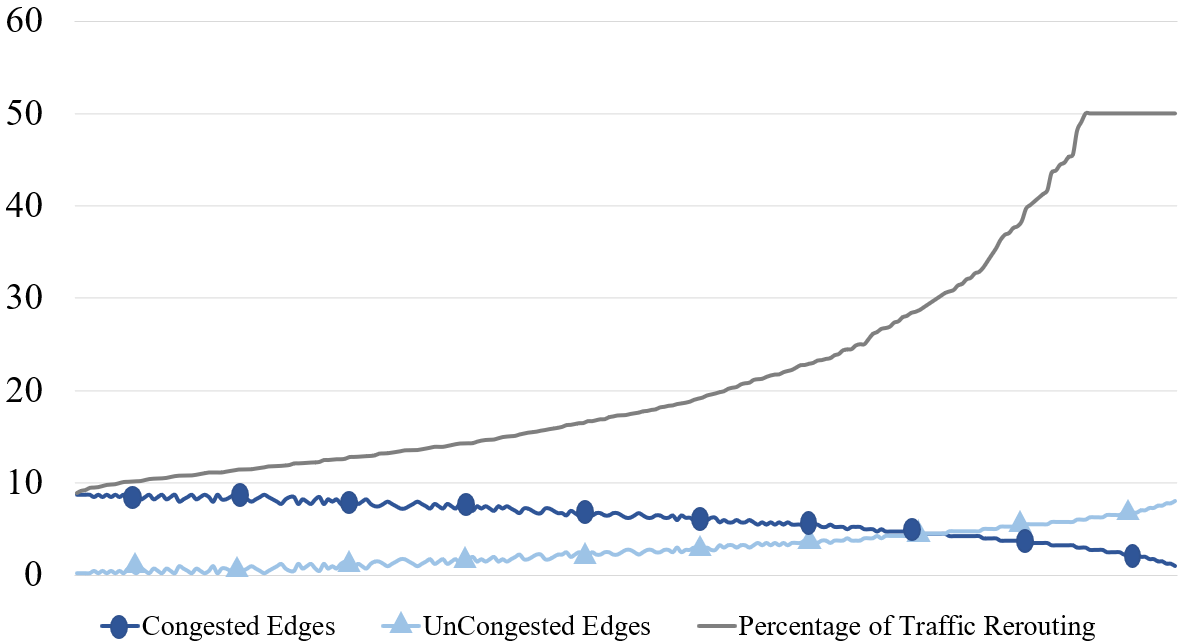}
    \caption{Percentage of network traffic rerouting with the number of congestion and uncongestion edges on the vertex.}
    \label{fig:TRG}
    \vspace{-1.5em}
\end{figure}

\[
\sum_{i=1}^{k} T_{sd_i} \leq P
\]

\begin{figure}
    \centering
    \includegraphics[width=1\linewidth]{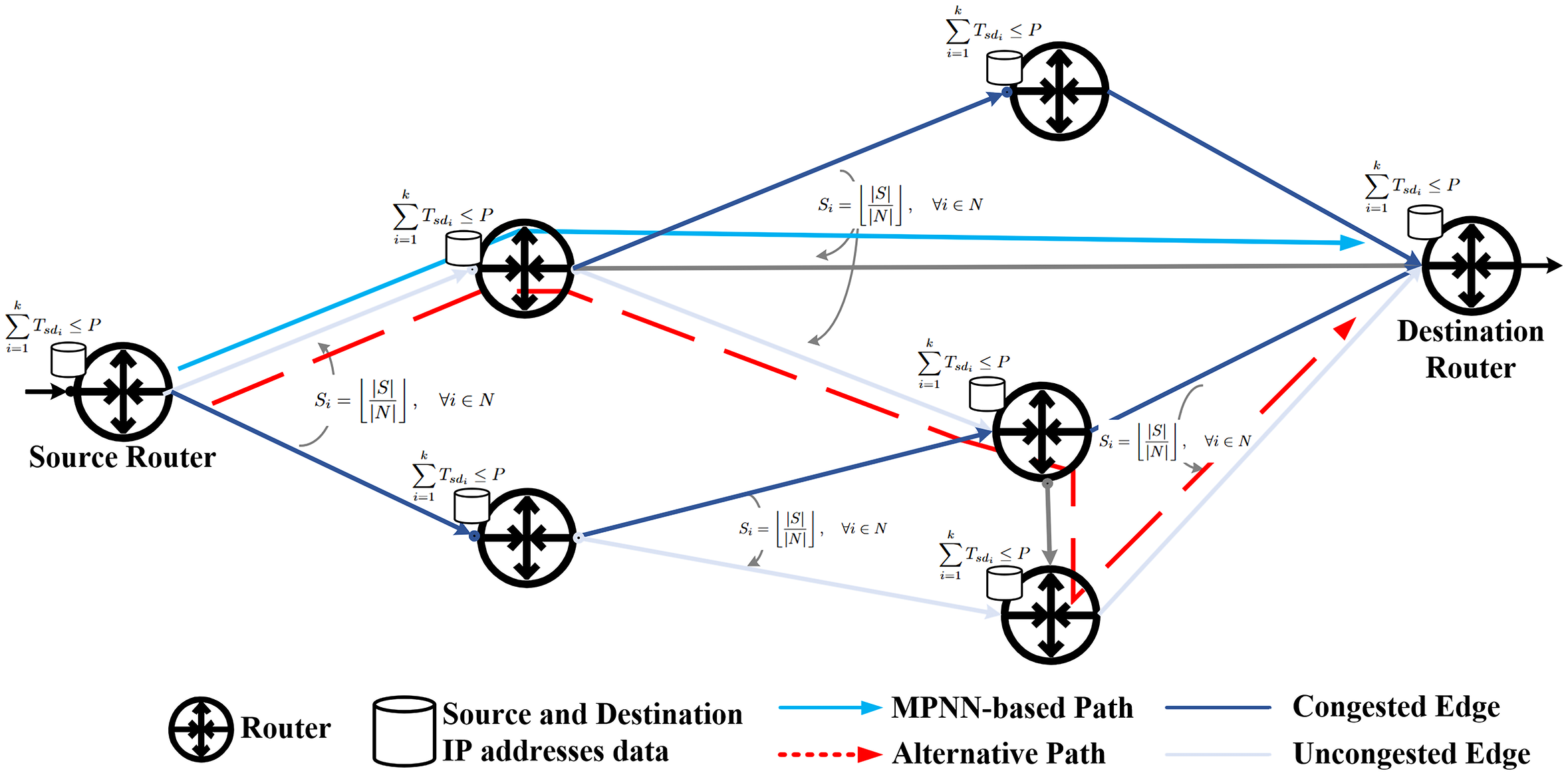}
      \caption{The router uses real-time traffic metrics to reroute traffic from congested to uncongested edges.}
    \label{fig:RS}
    \vspace{-2.0em}
\end{figure}

The selected S-D pairs are then redistributed across uncongested vertices as shown in fig. \ref{fig:RS}. Let, $S$ represents the set of rerouted S-D pairs and $E$ the set of uncongested edges. The number of S-D pairs assigned to each vertex $i$ is determined by:

\[
S_i = \left\lfloor \frac{|S|}{|E|} \right\rfloor, \quad \forall i \in E
\]

 This even distribution prevents overloading specific vertices and maintains overall network stability.

\section{Implementation}
\label{sec:implementation}

The \textbf{Bidirectional} communication in NDT has the objective  of  network optimisation and is carried out in a  controlled simulation environment using \textbf{GNS3} \cite{golightly2023deploying}.It uses a number of Python libraries, including \textbf{networkX}, \textbf{requests}, \textbf{telnetlib}, \textbf{rsync}, and \textbf{PyShark}. In the NDT, it  automates topologies generation using graph-based models using \textbf{networkx}, constructs the networks using the HTTP client library \textbf{Requests}, and configures them using \textbf{telnetlib}. Once network deployed, \textbf{rsync} used for controlled traffic generation to analyze the network behaviour, while \textbf{PyShark} retrieve packet data  from edges to extract key vertex and edge features. These extracted features, along with edge indices, are processed by a \textbf{MPNN} in \textbf{PyTorch} to classify congestion and support MPNN-driven PBR for dynamic network traffic management. The implementation is divided into six main parts: creating the topology, constructing the network, configuring it up, generating dynamic traffic, extracting features, and using MPNN to make routing more efficient for netown  traffic optimization.

The graph-based topologies generated by the Barabási Albert, Erdős Rényi, and Watts Strogatz models provide distinctive structural advantages while maintaining constrained vertex degree limitations (\(2 \leq \deg(v) \leq 9\)). The Erdős-Rényi model generates a topology by assigning edges based on the probability, \(p\), which makes sure that there are no isolated vertices. The Barabási–Albert model applies a preferential attachment mechanism, which enables vertices to connect according to their degree while imposing constraints to prevent over-connectivity. The Watts-Strogatz model begins with a structured ring lattice. Edges are mathematically rewired using probability p, introducing small-world features and degree limits for network stability and efficiency.

After the topology is generated, it is then deployed in \textbf{GNS3} and configured as an \textbf{MPLS (OSPF) network}, the network consists of \textbf{P (core) routers, PE (provider edge) routers, CE (customer edge) routers, switches, and hosts (Linux servers)}. CE routers are directly connected with PE routers, which makes it an MPLS backbone that facilitates efficient data transmission and routing. This topology deployment ensures network reliability under various operational conditions. \textbf{PE (Provider Edge (Boundary))} routers are represented as such \( V_{PE} \subseteq V \), these routers work as the gateway between the WAN and LAN. The formula below is used to calculate the number of PE routers with the P routers:
\[
V_{P_E} = \max\left(2, \frac{V_P}{5} + 1\right)
\]
\textbf{CE (Customer Edge)} routers are represented as such \( V_{CE} \subseteq V \), these routers interface directly with external customer networks. Each CE router connects one-to-one with a PE router. The number of CE routers is equal to the number of PE routers.
\textbf{Switches and Hosts (Linux Servers)}: Represented as \( S \subseteq V \)switches, they facilitate communication between CE routers and hosts \( H \subseteq V \). Each CE router connects to one switch, making the number of switches equal to the number of CE routers. Each switch connects to exactly three hosts, or we say that the number of hosts attached to the switches is three times as compared to PE routers.

To simulate real-world network behaviour, dynamic traffic flows are generated with the help of automated scripts that use \texttt{rsync} for controlled data transfers between host vertices. The traffic load increases  with an increase in file size in every iteration, which helps in assessing near real-time  performance for metrics  such as throughput, delay, and congestion for edges and vertices on the network. By subjecting the network to varying traffic loads, the topology efficiency can be evaluated to ensure the network remains adaptable under fluctuating conditions.

Additionally, there are a number of traffic generation approaches. It can be generated using \textbf{the ping command}, which is used to evaluate non-forwarding packets and path discovery and assess response patterns. By generating it with \textbf{NC (Netcat)}, it helps in testing the port and data transfer, but it has the overhead of developing logic for error handling. Another approach is \textbf{SCP,} which is used for secure file transfers, but it has the overhead of high CPU usage due to encryption. The approach we are using is \textbf{RSYNC}, which transfers files with built-in verification for synchronisation, optimised memory usage, and lower CPU utilisation. On every iteration old file are remove and new files are created with increase file size and transmitted to destination host through RSYNC unix utlity. Unlike Ping, SCP, and Netcat, RSYNC's built-in verification helps to avoid  data corruption, making it more reliable for sustained network load testing in high-traffic environments.

To capture meaningful performance metrics, packet-level data is analysed using PyShark. The packet frame provides the time \texttt{time\_delta} and packet length. The packet sniffing provides for packet arrival, previous reference, and departure  timestamps. The Ethernet header provides traffic flow direction by capturing source and destination MAC addresses, while the IP header provides source and destination IP addresses along with protocol details. This extracted information from the packet header helps to extract vertex and edge features such as \textbf{delay}, \textbf{throughput}, and \textbf{congestion}, along with edge indices, which are essential for network performance evaluation.

The extracted features and edge indices serve as input for the \textbf{MPNN-based congestion prediction model}. The \textbf{MPNN training pipeline} employs iterative \textbf{message passing} to propagate congestion states of edges attached to vertices in the network, capturing both local neighbouring vertex interactions and global traffic dynamics. The model classifies each edge into categories of \textbf{underutilised}, \textbf{balanced}, \textbf{moderately congested}, or \textbf{highly congested}. The system then computes a \textbf{traffic rerouting percentage} based on the ratio of congested and uncongested edges for each vertex, adjusting traffic distribution accordingly.

As network conditions evolve, the \textbf{Traffic Rerouting Percentage} is dynamically updated in the MPNN-based PBR. Adjustments are made by modifying \textbf{Access Control Lists (ACLs)}, which store source-destination IP pairs that require rerouting. Each ACL is linked to a route-map policy that directs traffic toward uncongested paths. To ensure seamless packet delivery, the system verifies reachability using the \textbf{ping command}, prevents routing loops with \textbf{traceroute}, and avoids rerouting traffic directly attached to PE vertices. Routing table entries further confirm successful data transmission.

Guided by the \textbf{MPNN}, the \textbf{PBR routing mechanism} continuously adapts to network congestion. By dynamically responding to traffic patterns, it minimises delay, optimises throughput, and maintains balanced routing. Underutilised vertices receive appropriate traffic, preventing congestion buildup. The effectiveness of this strategy is evaluated by comparing MPNN-based PBR with traditional PBR, leveraging real-time monitoring of \textbf{latency}, \textbf{throughput}, and \textbf{congestion} to optimise routing policies.

\section{Experiment}
\label{sec:experiment}

There are two phases in the experimental process: Phase One sets a baseline by making traffic without MPNN-based routing and gathering traffic data from edges to train the MPNN model. In Phase Two, the trained MPNN is used to dynamically optimise traffic routing by looking at changes in congestion, delay, and throughput.

\subsection{Phase One: Establishing the Baseline}
The first phase establishes the foundation by successively generating traffic across all topologies without utilising the Message Passing Neural Network (MPNN) model and rerouting it via MPLS (OSPF). Traffic is generated iteratively, with file sizes increasing gradually at every step. During this phase, packet data from each edge of the simulated network is collected, providing a dataset for training the MPNN model for classification edges on each vertex of network. This dataset has also been stored for future analysis and utilisation. The findings of this phase serve as a benchmark to evaluate the changes implemented in Phase Two.

\subsection{Phase Two: Implementing the MPNN Model}
After training, the MPNN model is employed as a module in NDT, with PBR feedback, to optimise traffic in each network topology. For network behaviours to be captured, use the same iterative traffic creation procedure as Phase One. However, during this phase, network behaviour is actively studied, with data retrieval in real time in one direction from the simulated network to its digital twin, where the MPNN model analyses collected network vertice and edge features. The MPNN model classifies the edges for each vertex and then generates MPNN-based PBR commands, which are subsequently sent back to the simulated network to reconfigure the routers and adjust traffic flows dynamically in response to real-time conditions.

This two-phase experimentation enables an evaluation of network features such as congestion,  throughput, delay, and file transfer rates. By comparing the results from both phases, we assess the impact of the MPNN-based PBR, determining whether network congestion is balanced, delay is reduced,  and throughput is enhanced. Through this experimentation, the NDT is validated as an adaptive solution for optimising network traffic rerouting under dynamic load and path variations.

\section{Results and Discussions}
\label{sec:results}

The NDT shows improvement by addressing challenges in managing dynamic traffic and adapting to changes in the physical network’s topology. It has established a feedback looping mechanism that evaluates gathered metrics, and it updates routing decisions through programmable protocol on each vertex, step by step transforming it into a fully programmable physical network. In order to gain this, the NDT has integrated practical graph constraints, enabling topologies that scale and adapt effectively. It has also strengthened programmable control and reduced reliance on physical testbeds by using virtualised simulations.  Continuously monitoring the physical network, the NDT extracts edge- and vertex-level metrics to send routing feedback using MPNN-based PBR to each vertex on the physical network. The feedback has resulted in optimise traffic flow  by handling the dynamic changes on network traffic and topology in near real-time . This adjustment is done by rerouting traffic away from congested edges, which improves the use of available edges and vertices. The improvement is assessed through a comparative analysis of MPNN-based PBR and MPLS (OSPF). The analysis shows that factors such as lower congestion on edges of the router, a shorter queue, and reduced delay on edges results by having traffic optimization by handling same traffic by more edges on each router of the network. With shorter queues, reduced delay and more active edges operating in parallel, file transfers speed up and computational cost decreases. These improvements address earlier challenges and strengthen scalability and adaptability by balancing network traffic load on telecom infrastructures to support future traffic growth and increasingly complex topologies.

\subsection{Vertices and Edges Utilization}
\begin{figure}
    \centering
    \includegraphics[width=1\linewidth]{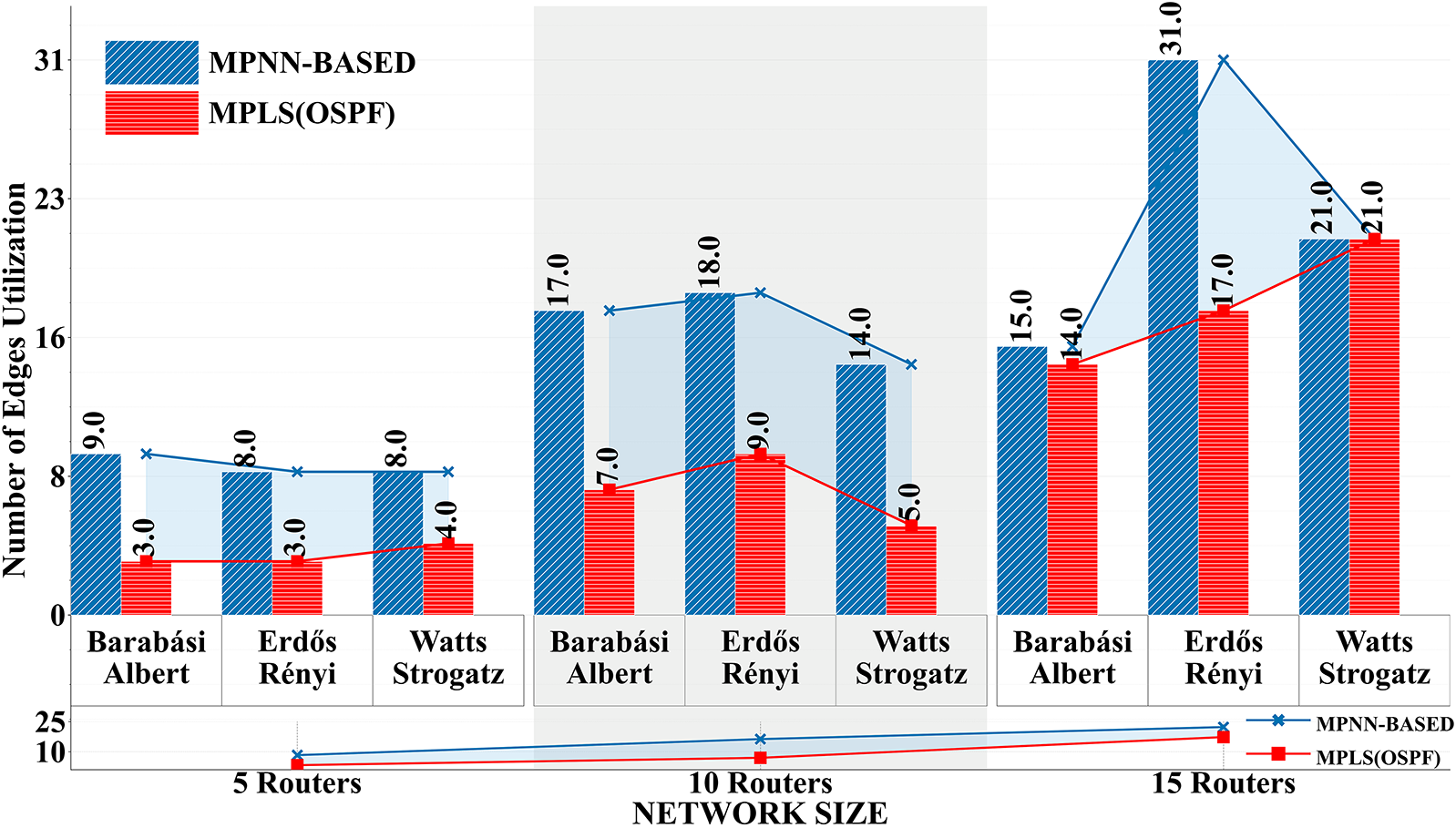}

                                                                \vspace{-0.5em}
 \caption{Network traffic transmitted by the MPNN-based solution demonstrates higher edge utilisation in comparison to MPLS (OSPF).}
    \label{fig:EC}
\vspace{-0.5em}
\end{figure}

\begin{figure}
    \centering
    \includegraphics[width=1\linewidth]{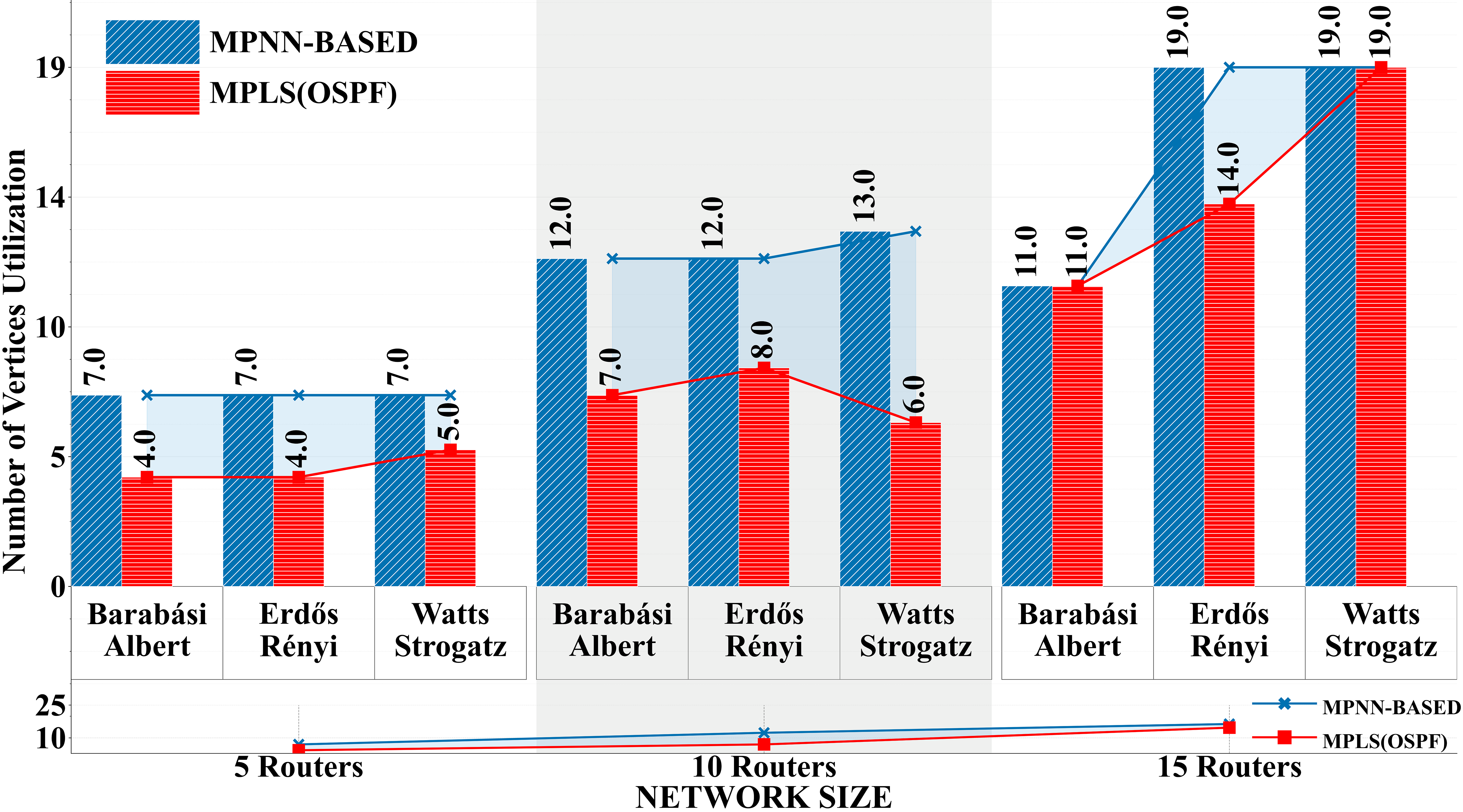}
                                \vspace{-1.5em}
 \caption{Network traffic transmitted by the MPNN-based solution demonstrates higher vertex utilisation in comparison to MPLS (OSPF).}
    \label{fig:NC}
       \vspace{-1.5em}
\end{figure}

Efficient use of vertices and links is essential for maintaining balanced traffic flow under dynamic changes in physical networks. For this NDT uses MPNN-based PBR routing to reroute the traffic on the router of the network, which results in increasing the utilisation of vertices and edges. The utilisation through rerouting is adaptable and scalable because NDT knows the local router and global network behaviour, which helps to the reroutes the traffic separately from congested to uncongested edges on each router of network. When the networks grow and new vertices are added, the share of active edges often drops, but the NDT keeps utilisation higher by adapting routing in near real-time, compared with routing protocol MPLS (OSPF). The vertex with a high degree steps in to handle heavier traffic and maintain stability, while lower-degree vertices assist where possible. Due to this nature of NDT that uses MPNN-based PBR, it achieved a 45.55\% increase in vertex utilisation as shown in fig. \ref{fig:NC} and an  88.94\% increase in edge utilisation as shown in fig. \ref{fig:EC}, compared with MPLS (OSPF), showing how it improves the traffic distribution of the network. Together, these actions help prevent bottlenecks, maintain smooth traffic movement, and strengthen scalability and adaptability so the network can grow without performance loss even as traffic and topology evolve.

\subsection{Impact of MPNN-Based PBR on Performance Metrices}
The different network topologies  with an increasing number of P routers are evaluated by transferring files across the network through routers, incrementally increasing the file sizes to analyse their effects on \textbf{delay, congestion, computational cost, and throughput} metrics for comparison between the MPNN-based PBR and MPLS (OSPF).

\subsubsection{Delay}
\begin{figure}
    \centering
    \includegraphics[width=1\linewidth]{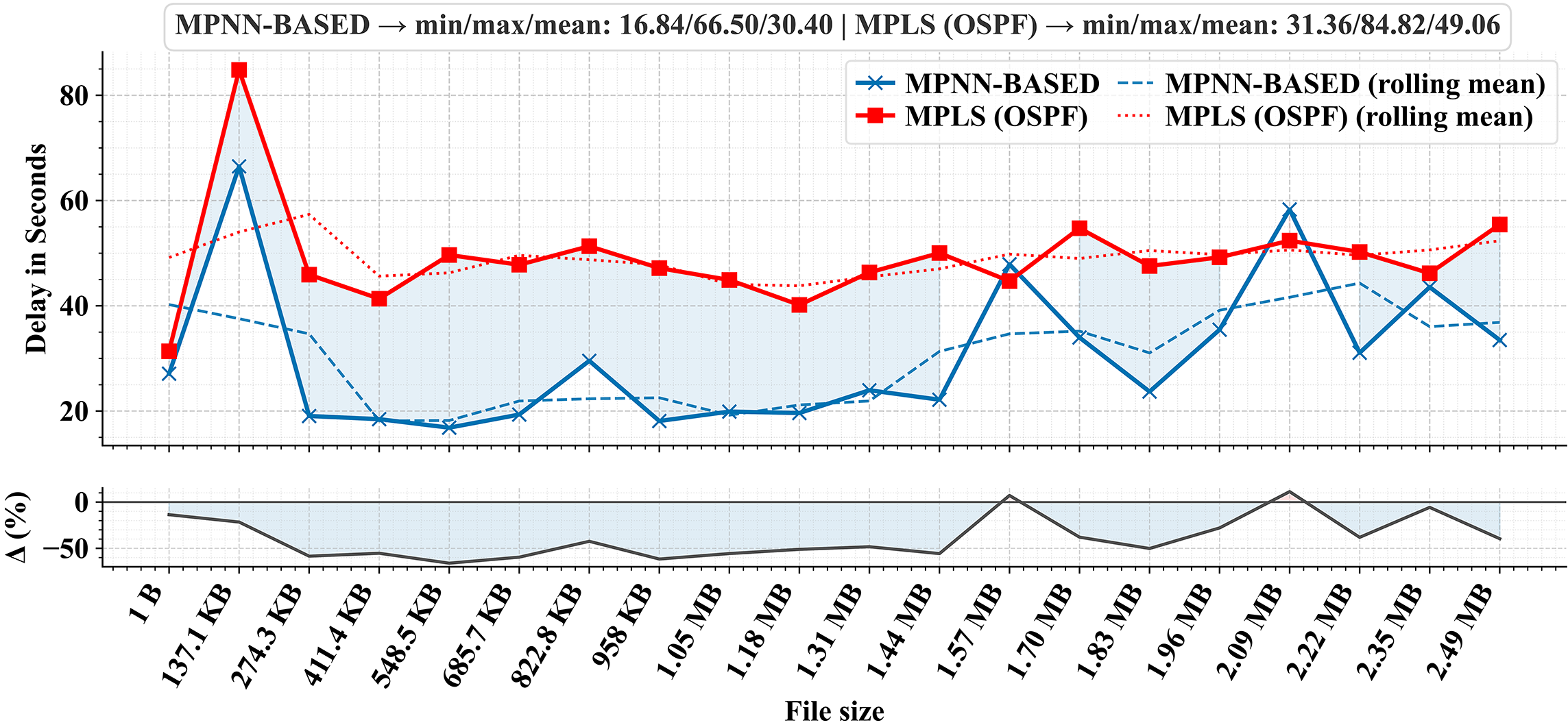} 
                                \vspace{-1.5em}
     \caption{End-to-end file transmission delays for varying file sizes, showing lower edge delays in MPNN-based compared to MPLS (OSPF).}
    \label{fig:D_F}
      \vspace{-0.5em}
\end{figure}

\begin{figure}
    \centering
    \includegraphics[width=1\linewidth]{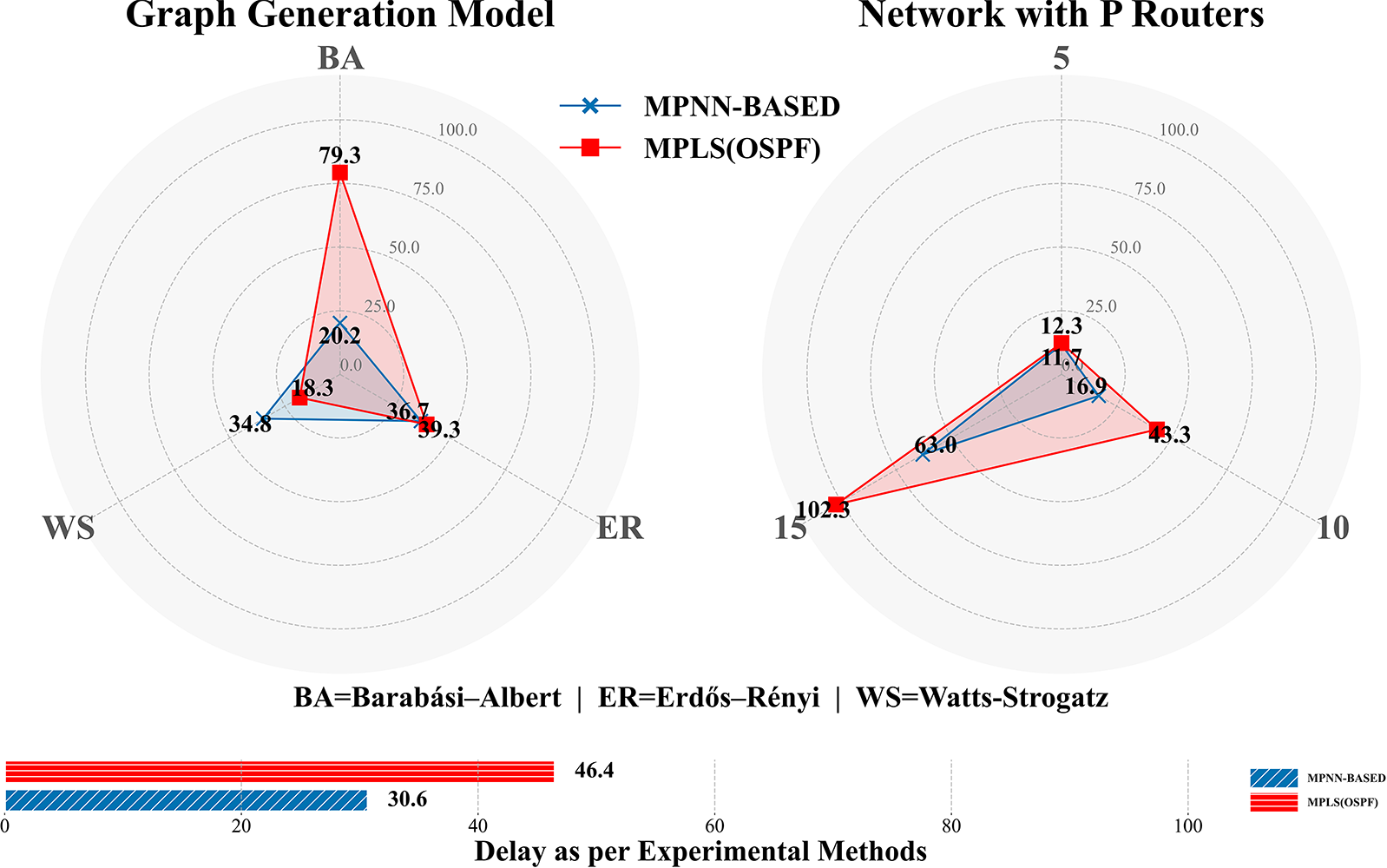}
                                                                                  
   \caption{Delay for overall network transmission based on graph generation models, P routers \& Experiments.}
     \label{fig:D_T}
       \vspace{-1.5em}
\end{figure}

Reducing delays has helped make traffic flow more smoothly in dynamic network conditions. The NDT shows demonstrable improvements over MPLS (OSPF) by intelligent routing that adapts to conditions in near real-time. When there is heavy traffic on the network, static MPLS (OSPF) routing frequently causes longer queues and higher delays. The NDT, on the other hand, reroutes traffic onto uncongested edges to balance traffic across the network and shorten queues, which results in a reduction of delay. This makes file transfers go faster and makes communication more reliable. The results of the experiments shown in  fig. \ref{fig:D_F} demonstrate that the NDT has reduced delay by 38.03\% as compared with MPLS (OSPF) under dynamically adjusting routing decisions. Topology-specific results in  fig. \ref{fig:D_T} demonstrate that Barabási–Albert has a 74.5\% delay reduction, Erdős–Rényi has a 6.7\% delay reduction, and Watts–Strogatz has a 89.9\% delay increase because it has a lower average vertex degree. Networks with 5, 10, and 15 P routers saw drops of 4.8\%, 60.8\%, and 38.4\%, respectively. The NDT reduces delay by 34.05\% on average compared to MPLS (OSPF). You can get these benefits through utilising parallel edges, making vertices more connected, and using different routing choices while avoiding OSPF convergence delays and label distribution costs and LSP setup, allowing for faster path selection \cite{al2024framework} and traffic forwarding without waiting for routing table population and label assignments.

\subsubsection{Congestion}

\begin{figure}
    \centering
    \includegraphics[width=1\linewidth]{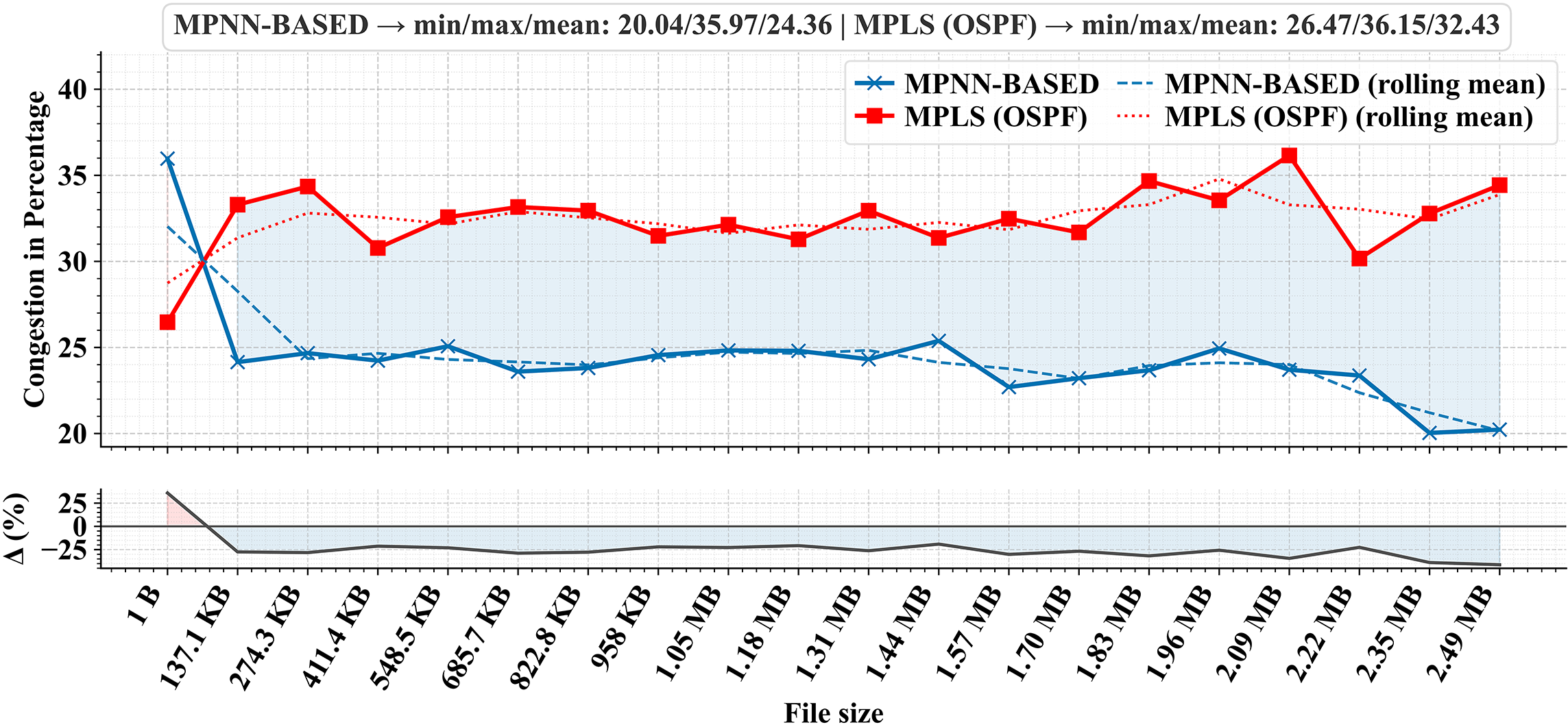}
                        \vspace{-1.5em}
      \caption{End-to-end file transmission congestion for varying file sizes, showing lower edge congestion in MPNN-based compared to MPLS (OSPF).}
    \label{fig:Cong_f}
   \vspace{-0.0em}
\end{figure}

\begin{figure}
    \centering
\vspace{-0.5em}
    \includegraphics[width=1\linewidth]{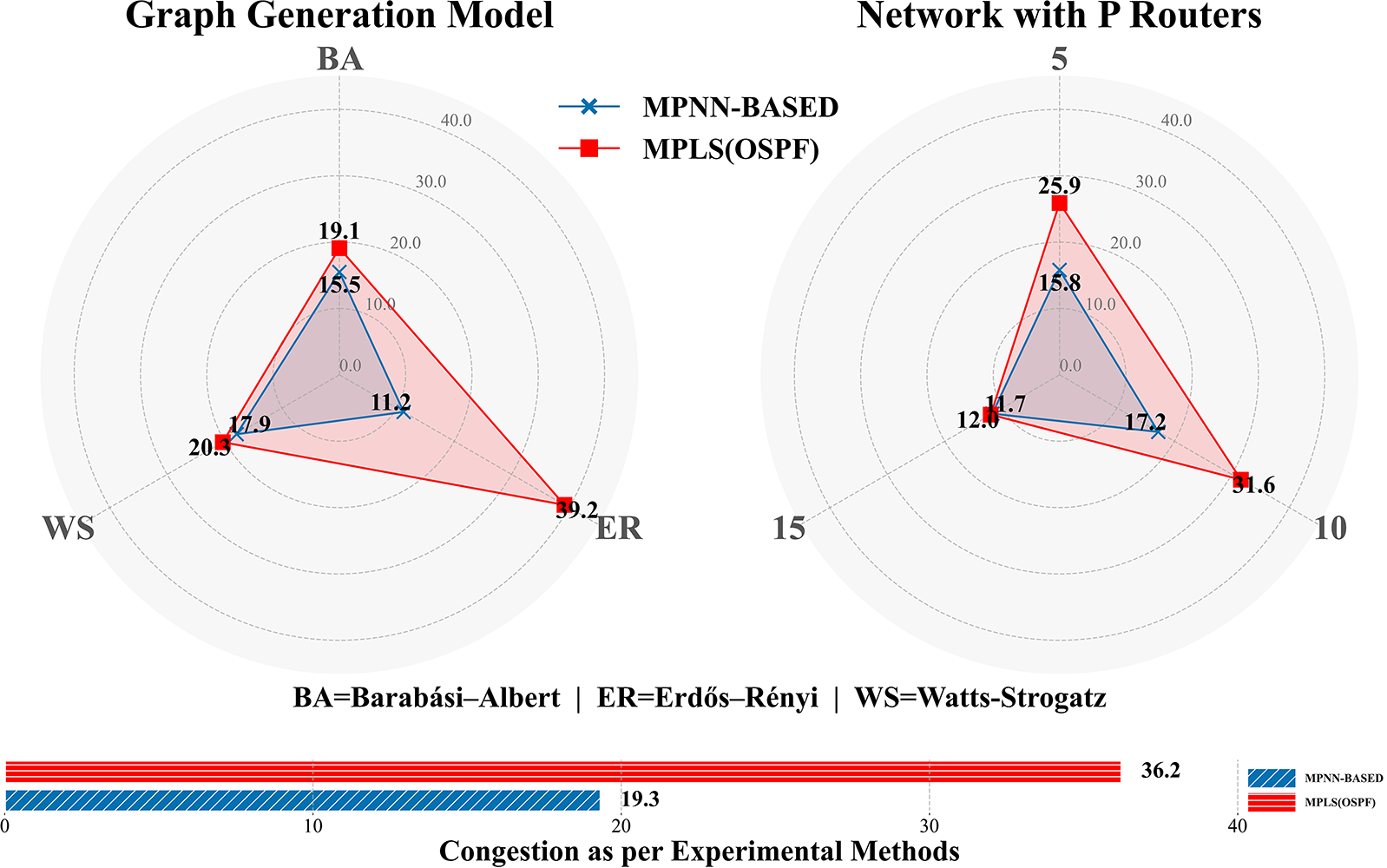}           
   \caption{Congestion for overall network transmission based on graph generation models, P routers \& Experiments}
    \label{fig:Cong_t}
 \vspace{-1.5em}
\end{figure}

As traffic and topology change, reducing congestion is essential for the network to perform optimally. The NDT addresses this problem by distributing traffic uniformly within the edges of each vertex. This keeps traffic moving smoothly and avoids places where it gets congested. It improves the approach for routing near real time by analysing both local vertex and global network behaviour to redirect traffic away from congested interfaces and keep the flow going smoothly. This approach shortens queue length, reduces congestion, and keeps throughput consistent with reduced delay at the edges of each vertex. Fig. \ref{fig:Cong_f}  shows that there is a 24.88\% decrease in congestion for end-to-end file transfers. Topology-specific results shown in fig. \ref{fig:Cong_t} that congestion reduces by 18.8\% in Barabási–Albert, 11.8\% in Watts–Strogatz and 71.4\% in Erdős–Rényi networks. Networks with 5, 10, and 15 P routers achieved congestion reductions of 38.9\%, 45.6\%, and 2.5\%, respectively. Compared to MPLS (OSPF), overall congestion gets reduced by 46.7\%. These results demonstrate that the NDT can adapt to changing network sizes, manage traffic in real time, and keep congestion reduced by balancing traffic across required edges for growing data-driven applications on the network.

\subsection{File Transfer Rate Improvement}

\begin{figure}
    \centering
    \includegraphics[width=1\linewidth]{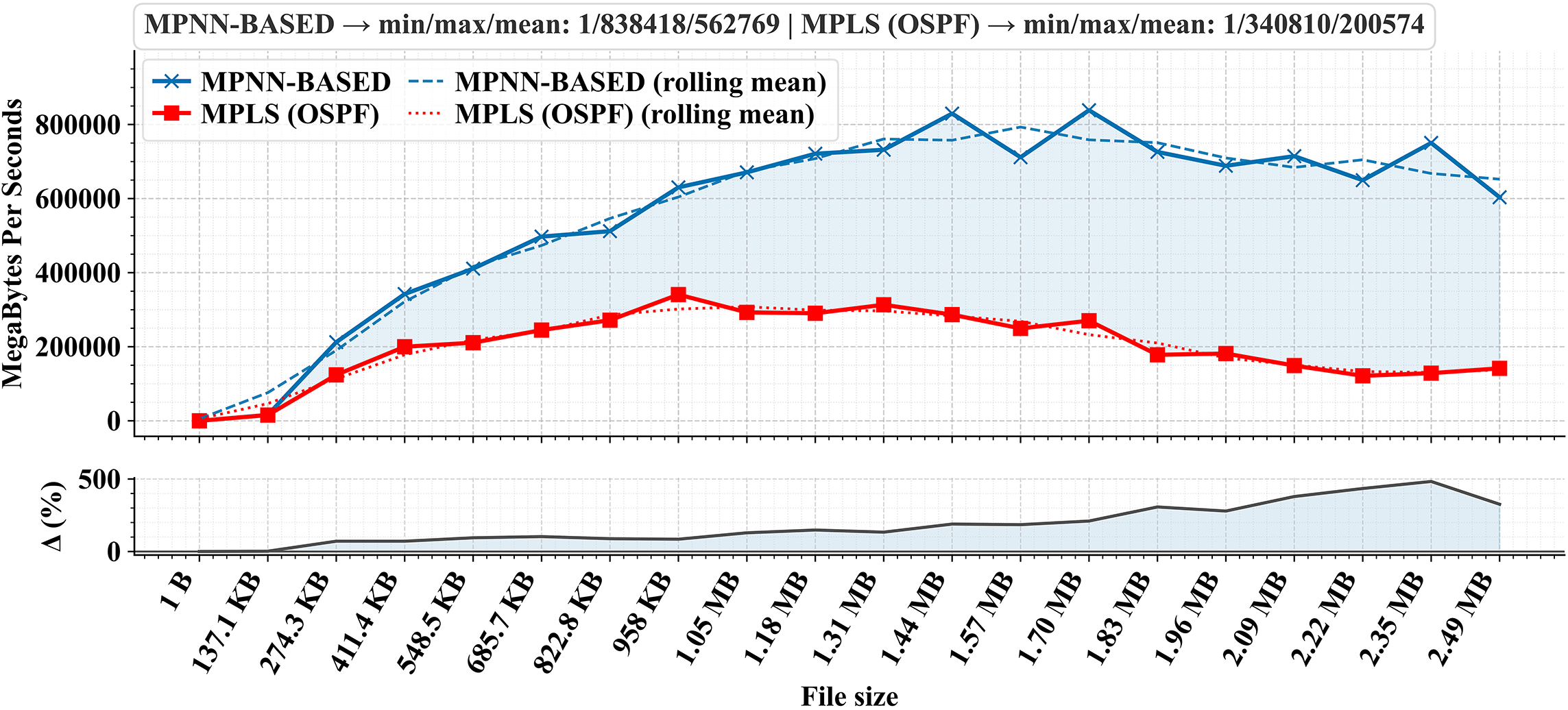}

                      \vspace{-0.5em}
    \caption{End-to-end file transfer rate for varying file sizes, showing a higher file transfer rate in MPNN-based compared to MPLS (OSPF).}
    \label{fig:frate_5}
     \vspace{-0.0em}
\end{figure}

\begin{figure}
    \centering
     \vspace{-0.5em}
    \includegraphics[width=1\linewidth]{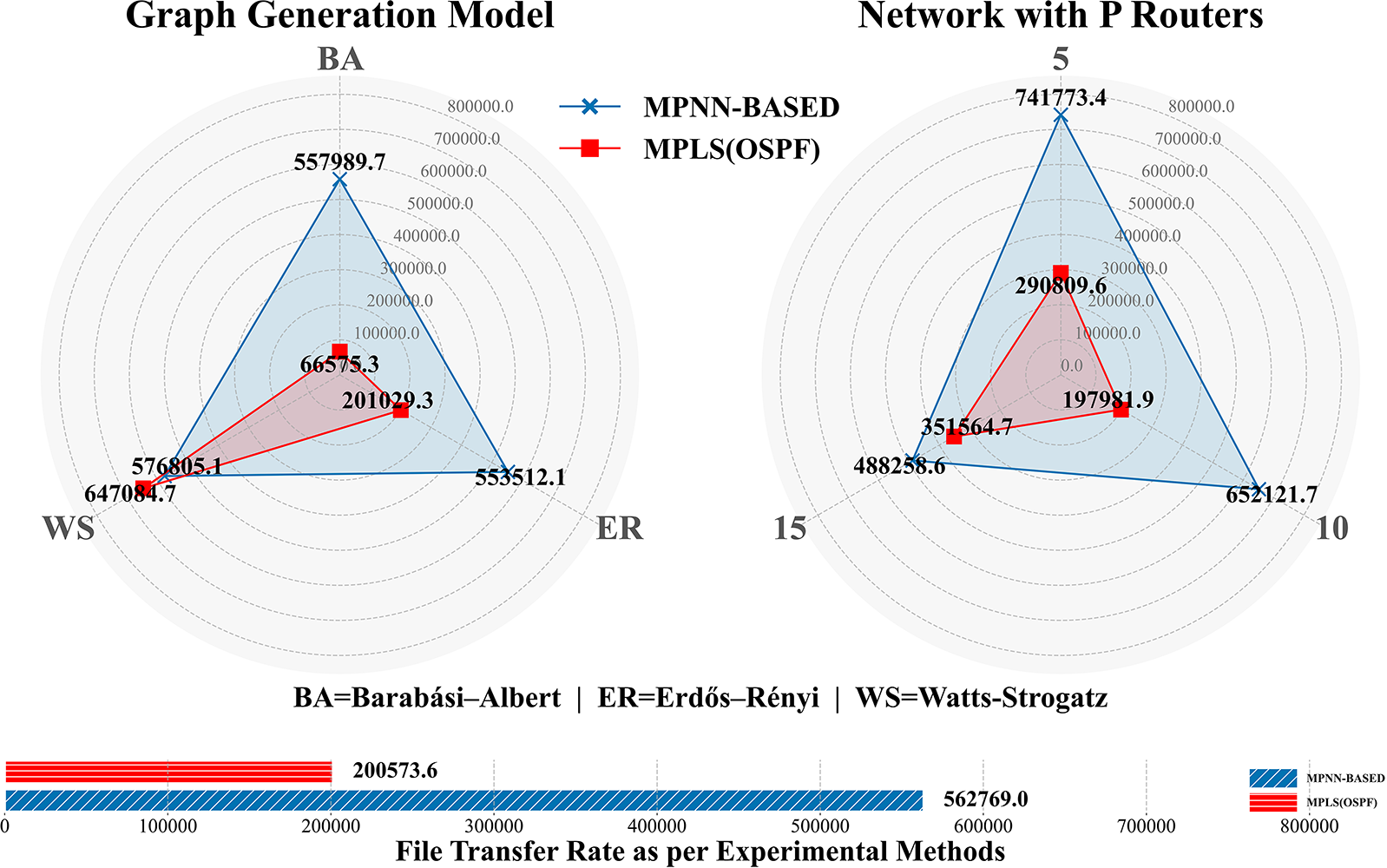}

      \caption{File transfer rate for overall network transmission based on graph generation models, P routers \& Experiments.}
    \label{fig:frate_10}
       \vspace{-1.5em}
\end{figure}

To ensure reliable end-to-end communication under dynamic network conditions, a faster file transfer rate is needed. The NDT improves file transfer rate by monitoring real-time traffic behaviour to reroute the traffic to uncongested edges of each router, which results in an increase in edges and vertices usage in the network under the dynamic network environment. As a result, file transfer performance improves, and networks become more resilient and adaptable. The NDT improves transfer rates over MPLS (OSPF) by interpreting real-time to make it more resilient and adaptable. This adaptability reduces transmission delays and sustains consistent speeds as networks evolve. Experiment results shown in fig. \ref{fig:frate_5} have increased by 180.6\% in file transfer rates compared with MPLS (OSPF). Topology-specific results in fig. \ref{fig:frate_10} indicate an 738.1\% gain in Barabási–Albert networks and 175.3\% in Erdős–Rényi, while Watts–Strogatz shows a 10.8\% decrease due to fewer alternative interfaces of routers that led to longer routes. Networks with 5, 10, and 15 P routers in size improve by 155.1\%, 229.4\%, and 38.9\%, respectively. Overall, the NDT achieves a 180.5\% higher transfer rate than other rerouting methods. These improvements prepare telecom networks to support increasing and more dynamic traffic with  faster delivery and performance for data-intensive applications.

\subsection{Computational Cost}
\begin{figure}
    \centering
    \includegraphics[width=1\linewidth]{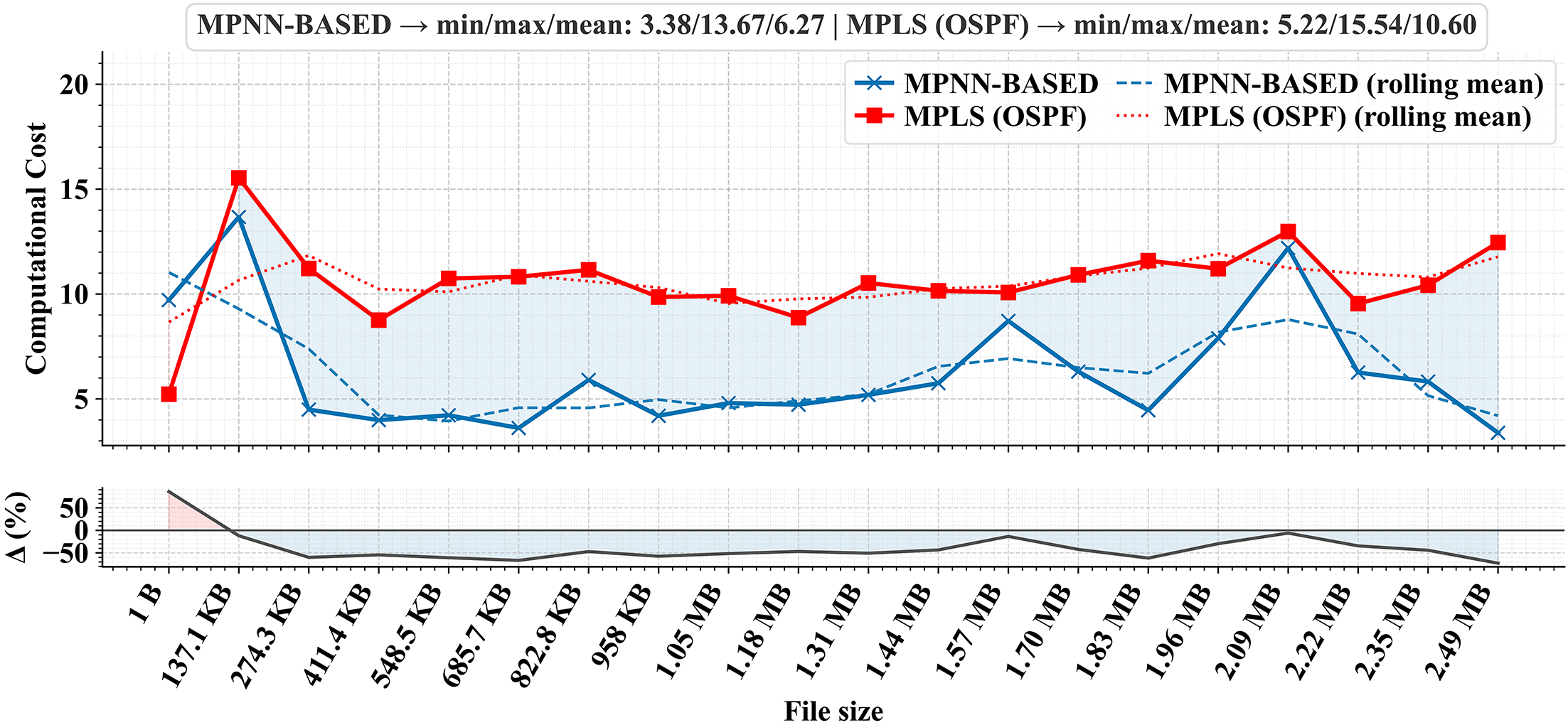}
     \vspace{-1.7em}
    \caption{End-to-end file transmission computational cost for varying file sizes, showing lower computational cost in MPNN-based compared to MPLS (OSPF).}
    \label{fig:CC_5}
     \vspace{-1.0em}
\end{figure}

\begin{figure}
    \centering
        
    \includegraphics[width=1\linewidth]{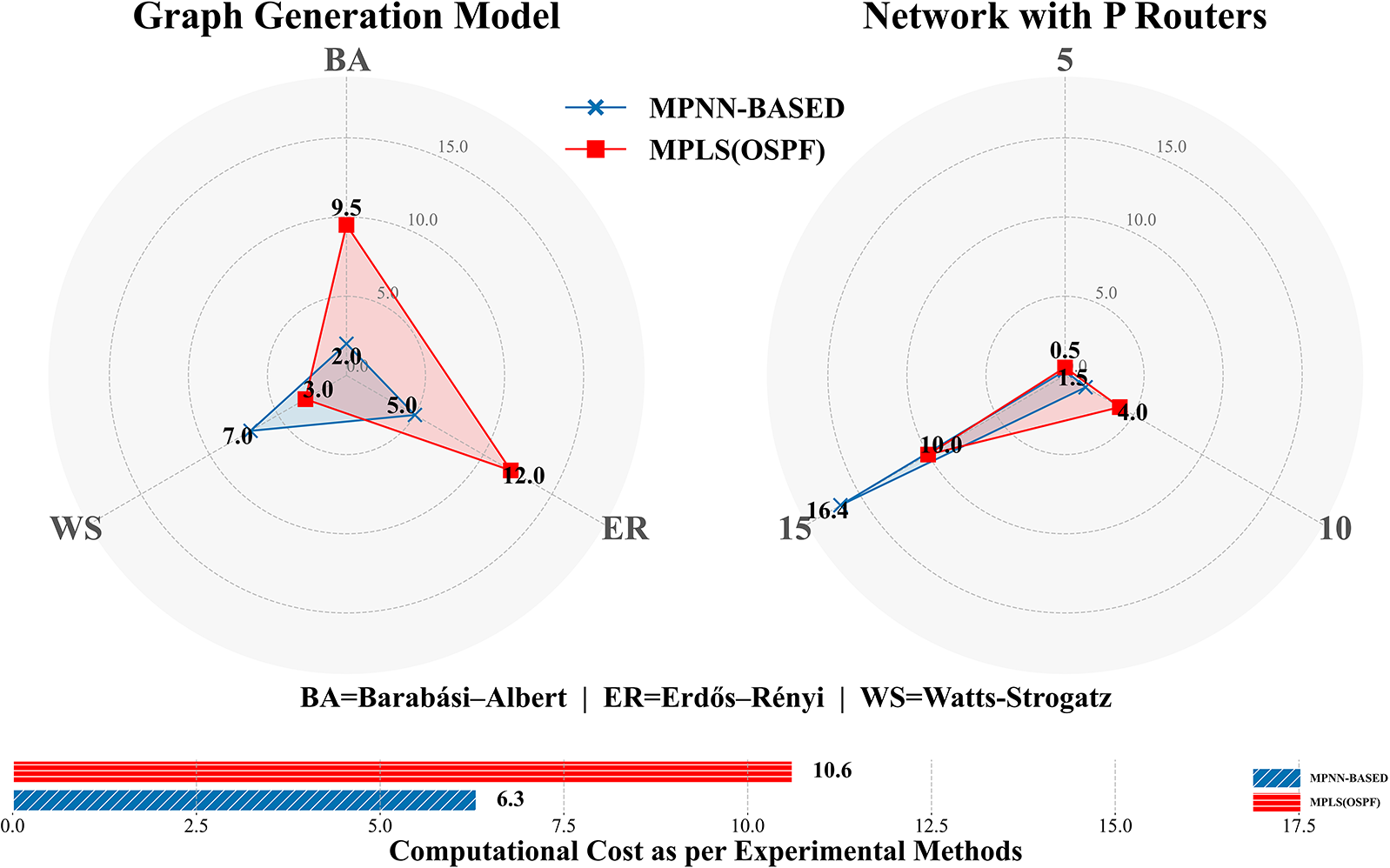}

  \caption{Computational cost for overall network transmission based on graph generation models, P routers \& Experiments.}
    \label{fig:CC_10}
       \vspace{-1.5em}
\end{figure}

Reducing computational cost is important for traffic flow optimisation as networks scale as well as changes in their topology. With MPNN-based PBR routing, NDT distributes traffic across more edges of each vertex. This reduces delay and queue length, which helps compute traffic faster. NDT also removes slow control-plane tasks like OSPF convergence, label distribution, and label-switched route creation, which cuts down on the amount of work that has to be done even further. So total processing overhead goes down, which lets vertices use fewer resources while still getting the job done. Fig. \ref{fig:CC_5} shows that the NDT reduces the cost of computing by an average of 40.88\% when transferring files of different sizes. In fig. \ref{fig:CC_10}, a more in-depth look indicates that the Barabási–Albert and Erdős–Rényi topologies save reducing by 79.1\% and 59\% on computational costs, whereas the Watts–Strogatz topology costs 137.9\% more since there are fewer potential pathways. Networks with 5, 10, and 15 P routers saw costs go down by 33.71\%, 60.79\% respectively. and network of 15 P routers cost goes by 64.41\% due to Watts–Strogatz topology. In general, the NDT costs up to 40.6\% less to run than other methods.   The dynamic recalculation of routing patterns at each file size increment causes this greater unpredictability. This changes traffic flows to minimise congestion and keep the network stable. The MPNN-based PBR cuts down on needless control-plane calculations by optimising edge use and spreading the processing burden evenly among routers. This improvement shows that it can effectively distribute resources, cut down on repetitive computations, and minimise the total cost of handling changing network circumstances.

\subsubsection{Throughput}

\begin{figure}
    \centering
    \includegraphics[width=1\linewidth]{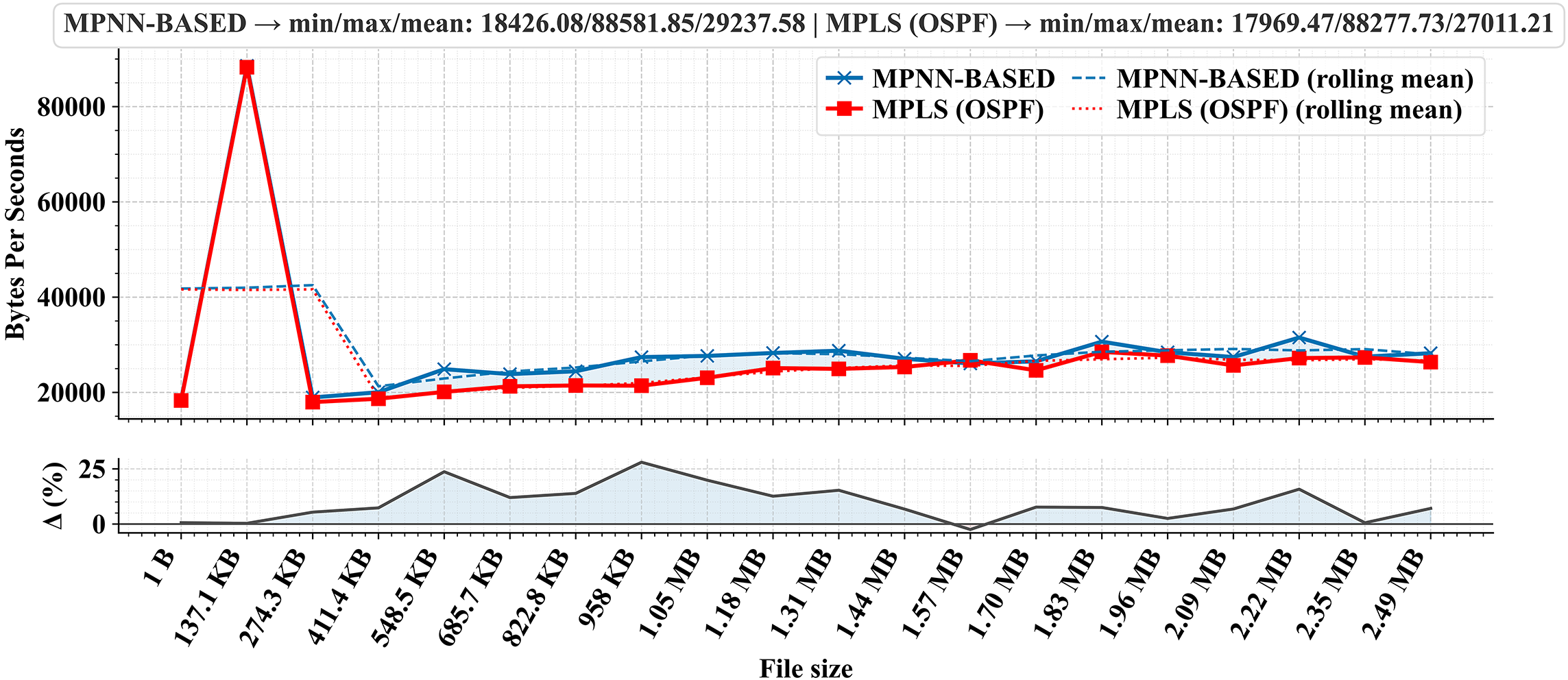}
                      \vspace{-1.5em}
      \caption{End-to-end file transmission throughput for varying file sizes, showing higher edge throughput in MPNN-based compared to MPLS (OSPF).}
    \label{fig:TP_F}
       \vspace{-0.5em}
\end{figure}

\begin{figure}
    \centering
    \includegraphics[width=1\linewidth]{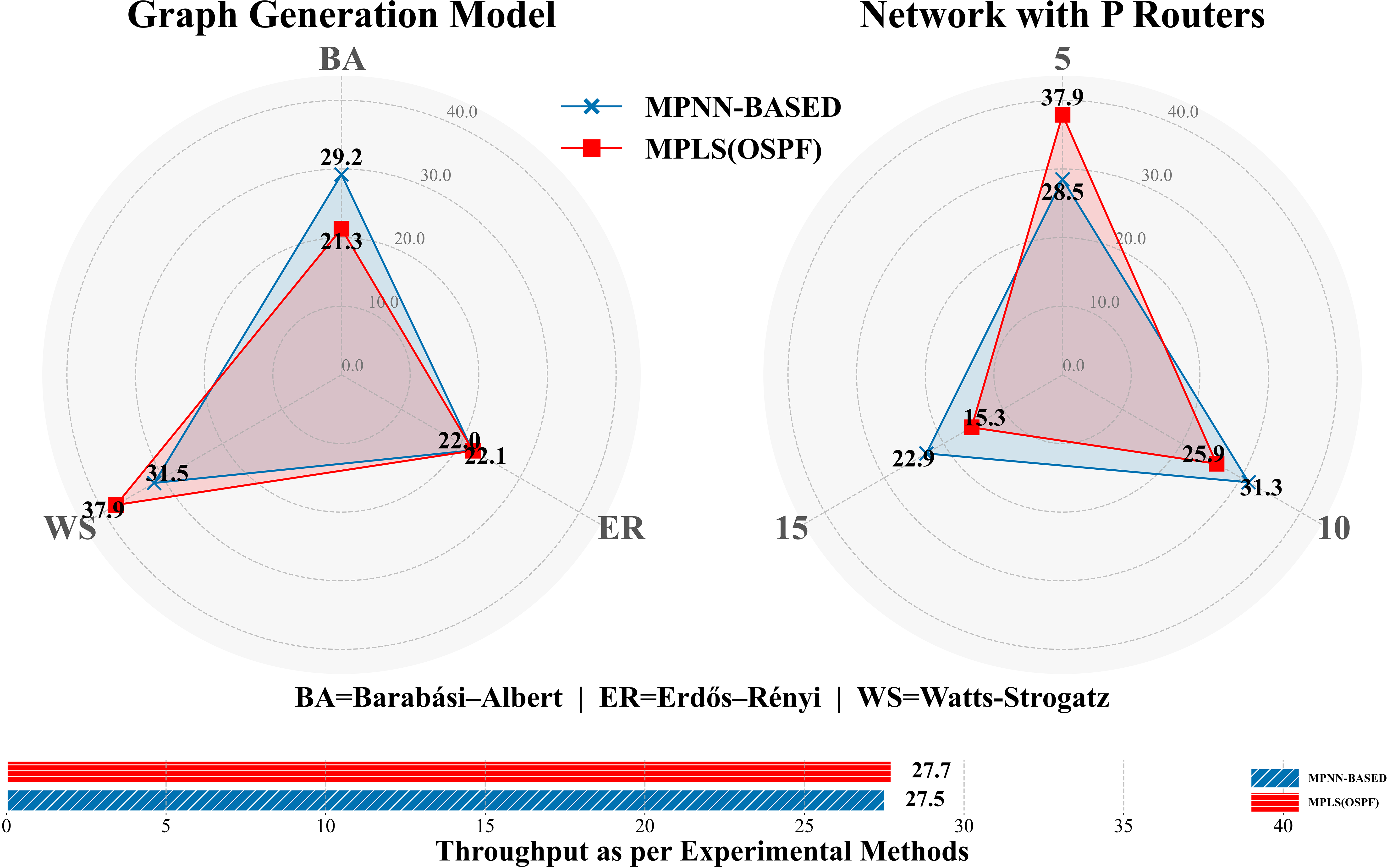}
                 \vspace{-1.5em}
       \caption{Throughput for overall network transmission based on graph generation models,, P routers \& Experiments}
   \label{fig:TP_T}
      \vspace{-1.5em}
\end{figure}

Throughput is important for making sure that networks can handle increasing traffic demands while maintaining reliable performance. The NDT maintains almost the same throughput as MPLS (OSPF) at the edges of the network because it sees congestion as a relative condition. An interface becomes congested when its throughput threshold is reached. This intelligent thresholding prevents packet loss and large queues from building up, which would slow down traffic flow. Unlike traditional static routing, which limits the number of interfaces and pathways and sometimes causes bottlenecks, the NDT adjusts in real time, rerouting traffic to required interfaces and paths to maintain performance as the network scales and evolves. The experiments in fig. \ref{fig:TP_F} show an average throughput increase of 8.2\% in end-to-end file transfer as compared to MPLS (OSPF). A topology-specific study in fig. \ref{fig:TP_T} shows that Barabási–Albert graphs have a 37.1\% increase, whereas Watts–Strogatz and Erdős–Rényi topologies have decreases of 16.9\% and 0.46\%, respectively. When scaling by the number of P routers, networks with 5 routers see an 24.8\% decrease, whereas networks with 10 and 15 routers see an increase of 20.8\% and 49.7\%, respectively. In general, the NDT maintains a balanced throughput, with just a minor 0.7\% decrease as compared to MPLS (OSPF). These findings show that the NDT can predict, adapt and scale, which means that future networks will be able to handle traffic by rerouting flows intelligently and maintaining network performance.

\subsection{Performance Evaluation}
These performance evaluations provided a clear narrative of improvement due to the NDT's intelligent traffic re-routing. As the results above demonstrate, delay was much reduced as the queues were shorter on the edges. This was because traffic was evenly distributed among them, which also increased the number of edges and vertices used on the actual network. By naturally reducing delay, congestion was also reduced, which let traffic flow smoothly and avoided bottlenecks. These improvements made file transfers faster, made time-sensitive applications work better, and made the network as a whole more responsive. Throughput remained constant because the NDT always selected reliable forwarding interfaces, even in complicated topologies like Watts–Strogatz or networks with just a few P routers. These improvements help to reduce computational cost, and also rerouting becomes simpler, with fewer recalculations and reduced label management overhead. Once you combine these results together, you get a concise and logical narrative about how the NDT works effectively for congestion-aware traffic optimisation in a physical network.

The performance evaluation includes the network topologies, each impacting the results in different manners. In Barabási–Albert networks, hub vertices that were connected closely and shorter routing pathways increased throughput, reduced delays, and reduced the cost of computing. On the other hand, Watts–Strogatz networks had more delays and a higher computational cost since they had fewer interfaces and longer routing pathways. Erdős-Rényi networks helped reduce congestion and computation costs by providing multiple route pathways. These research efforts show how topology-aware routing results in more balanced, efficient, and flexible telecom networks.
\section{Conclusion and Future Work}
\label{sec:conclusion}
In conclusion, this research offered an innovative network digital twin (NDT) that improves the performance of network traffic distribution in near real-time. It uses the message passing neural network(MPNN) to generate the policy based routing (PBR) strategy, which enhances network performance by \textbf{balancing throughput  and decreasing congestion, computational cost, and delays}. Unlike conventional MPLS, which may run across congestion in high-traffic areas, MPNN-based PBR dynamically redistributes traffic across available edges in near real-time to prevent any one edge from getting too crowded. Such an approach ensures a more efficient use of network resources and maintains optimal performance even under high traffic loads. The network digital twin helps to dynamically reroute traffic based on real-time insights. This network digital twin  ensures a resilient and efficient network capable of supporting next-generation applications like  \textbf{ B5G, IoT, and real-time applications}. Future research will expand this work by  using \textbf{multi-agent reinforcement learning (MARL) for further optimisation}, as well as applications in \textbf{scalability, service rerouting, energy cost optimization and network planning} to further enhance the resilience and efficiency of telecom networks.

\bibliographystyle{unsrt}
\bibliography{Reference}

\end{document}